\documentclass[journal,letterpaper,draftclsnofoot,onecolumn,12pt]{IEEEtran}

\usepackage{setspace}
\usepackage{cite}
\ifCLASSINFOpdf
\else
   \usepackage[dvips]{graphicx}
   \graphicspath{{../figures/}}
   \DeclareGraphicsExtensions{.eps}
\fi
\usepackage[cmex10]{amsmath}
\usepackage{amssymb,amsthm}
\usepackage{dsfont}

\usepackage{mathtools}

\newtheorem{remark}{Remark}

\newtheorem{theorem}{Theorem}
\newtheorem{corollary}{Corollary}
\newtheorem{definition}{Definition}
\newtheorem{identity}{Identity}

\newcommand{\sinr}{\mathtt{SINR}}
\newcommand{\op}{\mathtt{P}_{\text{\textnormal{o}}}}
\newcommand{\opt}{\mathtt{\tilde P}_{\text{\textnormal{o}}}}

\usepackage{psfrag}
\usepackage{array}
\usepackage{upgreek}

\usepackage{mdwmath}
\usepackage{mdwtab}

\usepackage{enumerate}

\usepackage{eqparbox}

\usepackage{mathtools}

\usepackage[caption=false,font=footnotesize,labelfont=sf,textfont=sf]{subfig}

\usepackage{stfloats}

\begin{document}
%\title{Interference in Aloha-based Ad Hoc Networks with\\Isotropic Node Distribution and Rayleigh Fading}
\title{Interference in Poisson Networks with Isotropically Distributed Nodes}

%\author{\IEEEauthorblockN{Ralph Tanbourgi, Holger J\"{a}kel, and Friedrich K. Jondral}
%\IEEEauthorblockA{Communications Engineering Lab, Karlsruhe Institute of Technology, Germany\\
%Email: \{ralph.tanbourgi, holger.jaekel, friedrich.jondral\}@kit.edu}
%}

\author{Ralph Tanbourgi, \IEEEmembership{Student Member, IEEE}, Holger J\"{a}kel, \IEEEmembership{Member, IEEE} and Friedrich K. Jondral, \IEEEmembership{Senior Member, IEEE}
%\thanks{Manuscript received January 20, 2002; revise
%d January 30, 2002. This work was supported by the I
%EEE.}%
\thanks{The authors are with the Communications Engineering Lab, Karlsruhe Institute of Technology, Germany. Email: \texttt{\{ralph.tanbourgi, holger.jaekel, friedrich.jondral\}@kit.edu}. The authors gratefully acknowledge that their work is partially supported within the Priority Program 1397 "COIN" under grant No. JO 258/21-1 by the German Research Foundation (DFG). This work was presented in part at IEEE Int. Symposium on Inf. Theory, MA, USA, July 2012.}}

\maketitle

\renewcommand\qedsymbol{\IEEEQEDopen}
\newcommand{\widesim}[2][1.5]{%
  \mathrel{\overset{#2}{\scalebox{#1}[1]{\raisebox{-0.08cm}{$\sim$}}}}
}

\begin{abstract}
%\boldmath
Practical wireless networks are finite, and hence non-stationary with nodes typically non-homo-geneously deployed over the area. This leads to a location-dependent performance and to boundary effects which are both often neglected in network modeling. In this work, interference in networks with nodes distributed according to an isotropic but not necessarily stationary Poisson point process (PPP) are studied. The resulting link performance is precisely characterized as a function of (i) an arbitrary receiver location and of (ii) an arbitrary isotropic shape of the spatial distribution. Closed-form expressions for the first moment and the Laplace transform of the interference are derived for the path loss exponents $\alpha=2$ and $\alpha=4$, and simple bounds are derived for other cases. The developed model is applied to practical problems in network analysis: for instance, the accuracy loss due to neglecting border effects is shown to be undesirably high within transition regions of certain deployment scenarios. Using a throughput metric not relying on the stationarity of the spatial node distribution, the spatial throughput locally around a given node is characterized.  

%The interference distribution is characterized through a first moment analysis for arbitrary block-fading channels (including the pure path loss model) and bounds on the tail probability are derived. For Rayleigh fading, the Laplace transform of the interference distribution is presented. For the path losses $\alpha=2$ and $\alpha=4$ closed-form results are derived. The developed theory is used to address 

%Two metrics suitable for measuring local throughput in non-stationary networks are proposed, and they are discussed for the isotropic model at hand. Furthermore, , particularly it revises some prior results for the case $\alpha=2$. On the other, it provides a powerful tool for studying non-stationary networks as demonstrated through numerous examples.

\end{abstract}


% \fi
\IEEEpeerreviewmaketitle

%\footnote{The authors gratefully acknowledge that their work is partially
%supported within the priority program 1397 "COIN" under grant No. JO
%258/21-1 by the German Research Foundation (DFG).}
%\footnote{This work is partially supported within the priority program 1397 "COIN" under grant No. JO
%258/21-1 by the German Research Foundation (DFG).}
\section{Introduction}\label{sec:introduction}

\textit{Stochastic geometry}, in particular the theory of point processes, has recently attracted much attention in the field of interference modeling and performance analysis for wireless networks with many uncertainties such as mobile/dynamic user locations and channel fading. %Originally inspired by statistical problems in material sciences \cite{stoyan95}, it to answer some of the fundamental questions in wireless network theory; thereby bridging the gap between yet unsolved multi-user communication problems in information theory and design guidelines for network operators. %Among the numerous contributions in this area the \emph{transmission capacity} framework led to many profound and practical results for wireless networks, cf. \cite{andrews10,weber10,weber05,andrews07,weber07,ganti09}.
% In essence, the transmission capacity framework models the spatial distribution of nodes of the network as being random rather than assuming a deterministic or fixed spatial configuration. The advantages hereof are manifold, see e.g. \cite{weber05,andrews10} for further details; but perhaps most significantly, such a probabilistic approach decouples the network model from the \emph{actual} spatial configuration which, in turn, increases the generality of obtainable results. Moreover, these results can be given in closed-form, thereby revealing the individual effects of the system parameters on the network performance. By definition, the transmission capacity gives the maximum density of successful transmissions that can take place simultaneously in the network, weighted by the probability of success of a typical transmission \cite{weber05}. 
In a nutshell, the locations of the nodes are modeled as a realization of a stochastic point process rather than assuming a fixed spatial configuration. Since the emitted signals undergo a distant-dependent path loss, the interference experienced by a given node becomes random. Its statistical properties, moreover, depend on several factors including the law of the \emph{spatial distribution} of nodes.

\subsection{Related Work and Motivation}
Interference modeling and network analysis using tools from stochastic geometry have become a multi-faceted research field \cite{sousa92,Ilow98,baccelli03,baccelli06,weber10,andrews10,weber05,andrews07,weber07,ganti09_clustered,baccelli09a,baccelli09b,ganti09,giacomelli11,ganti11,govindasamy11,blas12,hunter10,tanbourgi12,tanbourgi11_2}. Early works on interference modeling assumed a stationary PPP for the interferer locations, cf. \cite{sousa92,Ilow98}. Using a similar model, the spatial throughput of decentralized networks with Aloha medium access control (MAC) was analyzed in \cite{baccelli03,baccelli06}. Following these works, the node locations have mostly been modeled by \emph{stationary} point processes which typically leads to \emph{location-independent} statistical properties of the considered performance quantities, e.g., interference, outage probability or throughput. Among these advances, the transmission capacity framework \cite{weber10,andrews10,weber05,weber07,andrews07} substantially contributed to a better understanding of the interactions between the basic system parameters of a wireless network. Besides, stationary models with \emph{non-homogeneous} node deployments, e.g., Poisson-Cluster \cite{ganti09_clustered} and Mat\'{e}rn hard-core models \cite{baccelli09a,baccelli09b}, were also investigated as such models are well-suited for studying more sophisticated medium access control (MAC) schemes. Treated as \emph{general motion-invariant}, these models were further analyzed in \cite{ganti11,ganti09,giacomelli11} in a unifying way.
% Finite networks may be modeled by stationary cluster point processes and letting the \emph{parent density} tend to zero (with the representative cluster left). However, the problem of infinite interference for $\alpha=2$ is not solved by that as with all stationary models.

Stationarity of the spatial distribution of nodes is a desirable property since it allows for analytic tractability and, even more important, it represents a key requirement for applying certain performance metrics such as the transmission capacity metric \cite{weber10}. In practice though, wireless networks exhibit a non-stationary spatial node distribution; for instance, because of a finite network area with boundary regions. Consequently, performance-relevant quantities such as the experienced interference typically vary across the network area, thereby complicating modeling and system design. Besides this simple example, more complex deployments are often found in practice, e.g., wireless sensor networks created by airdrop \cite{akyildiz02} or spontaneous formation of hot spots \cite{feeney01}. The spatial configuration of such hotspots is typically dictated by user motion and by geometric constraints as illustrated in \cite{mit05} for the example of a campus-wide Wi-Fi deployment. Furthermore, there is a growing need for cellular operators to better understand not only the temporal variations in user traffic demands but also its spatial dependence; for instance, the optimal interplay between small-cell deployments and Wi-Fi offloading--a promising approach to boost capacity in dense areas--requires carefully pinpointing areas of peak-traffic demands \cite{mobidia12} and identifying locations for deployment so to reach the mobile users \cite{ruckus12}. Hence, analytic tools for quantifying the network performance while taking into account user mobility as well as hard-to-predict spatial configurations are of eminent importance.

The need for non-stationary models for characterizing more complex node deployments was reported for instance in \cite{avidor01}. The authors discussed techniques that generate non-uniform node distributions for the purpose of efficient network simulations. In \cite{govindasamy11}, a non-stationary and isotropic node distribution was assumed for analyzing multi-antenna receivers in the presence of interference. The analysis showed that the shape of the spatial distribution has a considerable impact on link performance. In \cite{tanbourgi12_3}, a first attempt was made towards analyzing the link performance at an arbitrary receiver location and for an arbitrary isotropic node distribution.

The shortcomings associated with the stationarity assumption are summarized below:

%\subsection{Limitations of the stationarity assumption}\label{sec:intro_limitations}
\textit{Infiniteness and boundary effects:} Stationarity implies that the network consists of infinitely many nodes spread over an infinitely large region. However, the number of nodes as well as the network area is finite in practice. Boundary effects are ignored although they play a critical role in real-world networks because of unequal performance among the nodes in terms of, e.g., local topology, interference-/noise-limited performance, etc. Ensuring the quality-of-service (QoS) level targeted before deployment hence becomes difficult.

\textit{Model artifacts for path loss exponent $\alpha=2$:} In the stationary case and with a path loss exponent of $\alpha=2$, the sum interference is almost surely (a.s.) infinite \cite{weber10}. %This model artifact results in an outage probability of one or, equivalently, to a transmission capacity equal to zero. 
More specifically, stationary models lose their accuracy as the path loss exponent $\alpha$ tends to $2$ since interference then becomes dominated by the infinite number of far nodes.

\textit{Lack of local throughput metric:} Non-stationary models prevent the use of certain throughput metrics such as the transmission capacity. This is because it is no longer possible to infer the global performance from the local analysis as the local performance is location-dependent in non-stationary deployments.%  When the node distribution is non-stationary, this metric must be modified to take into account heterogeneous node deployments.

\subsection{Contributions and Outcomes}
We extend prior work by modeling the node locations by an \textit{isotropic} PPP, with stationarity being a special case. The network model is explained in detail in Section~\ref{sec:model}. The contributions are summarized below.

\textit{Interference characterization:} The first moment and distribution function are studied for both arbitrary and Rayleigh fading channels. In the case of arbitrary fading, we derive in closed-form the exact first moment as a function of the spatial shape of the node distribution and of the arbitrary receiver location for the path loss exponents $\alpha=2$ and $\alpha=4$ in Section~\ref{sec:int_general_fading}. Using these results, an upper bound on the tail probability of the interference is derived. A corresponding lower bound that is not limited to the above values of $\alpha$ is also presented for suitable spatial shapes. In Section~\ref{sec:int_rayleigh_fading}, we derive the Laplace transform of the interference for the Rayleigh fading model. We also address the case $\alpha=2$, which eluded a meaningful analysis in the stationary model. One important insight is that for $\alpha=2$, one can find situations in which a.s. infinitely many nodes contribute to the interference while the interference remains finite a.s. This result sharpens prior statements about the nature of the interference for $\alpha=2$ and suggests that there exists a transition between sparse and dense networks.

\textit{Outage probability and model accuracy:} The location-dependent outage probability is characterized in Section~\ref{sec:op} as a function of the spatial shape. We demonstrate the use of the developed model in Section~\ref{sec:accuracy} by comparing it to a stationary model that uses a local approximation capturing the non-homogeneity in the spatial nodes distribution. We show that, depending on the spatial shape in question, large approximation errors are avoided by the developed model, particularly in transition areas where boundary effects come into play.

\textit{Applications:} We propose a metric that is capable of quantifying the local throughput in non-stationary networks in Section~\ref{sec:local_throughput} and demonstrate through an example how this metric can be used. In this example, we refine an existing result on code division multiple access (CDMA) systems in decentralized networks for the case $\alpha=2$. We also show how the model accuracy for carrier-sense based MACs in decentralized networks can be significantly increased by means of the developed model.

\textbf{Notation:} Sans-serif-style letters ($\mathsf{z}$) represent random variables while serif-style letters ($z$) represent deterministic variables or constants. %We denote by $\mathbb{P}\left(\cdot\right)$ and $\mathbb{E}\left[\cdot\right]$ the probability measure and the expectation operator, respectively. We will sometimes include the random variable under consideration in the subscript whenever necessary, e.g., $\mathbb{P}_{\mathsf{z}}\left(\cdot\right)$ and $\mathbb{E}_{\mathsf{z}}\left[\cdot\right]$. 
The imaginary unit is given by $j=\sqrt{-1}$ and $\mathfrak{R}(\cdot)$ denotes the real part of a complex-valued number. We define by $b(z,r)$ and $\partial b(z,r)$ a disc and a circle, respectively, centered at $z\in\mathbb{R}^2$ with radius $r>0$. The origin is denoted by $o$. 

\section{Mathematical Model}\label{sec:model}

We consider a packet-based wireless network with identically-equipped nodes isotropically distributed in $\mathbb{R}^2$. The nodes are assumed to be slot-synchronized. In a randomly chosen slot, some nodes wish to transmit a packet. We assume that the locations $\mathsf{x}_{1},\mathsf{x}_{2},\ldots$ of these transmitters follow an isotropic PPP $\Phi\triangleq\{\mathsf{x}_{i}\}_{i=1}^{\infty}$ on $\mathbb{R}^2$ with intensity (equivalently, density) $\lambda(x)$ being defined on $\mathbb{R}^2$. Throughout this work we will denote by $\mathsf{x}_{i}$ to the random location of the $i$-th node as well as to the $i$-th node itself. Due to isotropy of $\Phi$, $\lambda(x)$ is rotation-invariant \cite{stoyan95} and depends solely on the distance $\|x\|$ to the origin, i.e., $\lambda(x)=\lambda(\|x\|e^{j\theta})$ for all $\theta\in[-\pi,\pi)$.\footnote{In a very few situations, we will switch between Cartesian coordinates ($x$) and the corresponding polar coordinates ($|x|e^{j\theta}$) when appropriate. We do not expect any confusion thereof.} For notational convenience, we define $r\triangleq\|x\|$. The next definition is a consequence of the fact that $\lambda(x)$ can be described as the resulting intensity after \emph{location-dependent} thinning of a stationary PPP of some constant intensity \cite{baccelli09a}.
\begin{definition}\label{def:shape_function}
The shape function $F:\mathbb{R}_{\geq0}\mapsto[0,1]$, reflecting the \emph{spatial shape} of $\Phi$, is defined on by the relation $\lambda\left(x\right)=\lambda F(\|x\|)$, where %$F(r)\geq0$ for all $r\geq0$ and $\max_{r}\{F(r)\}=1$, and 
$0<\lambda<\infty$ is some intensity scaling constant.
\end{definition}
%The restrictions are necessary to ensure that $\lambda(x)$ is non-negative and bounded by $\lambda$ everywhere. The shape function $F(r)$ is defined on $\mathbb{R}_{\geq0}$. 
We assume that each transmitter $\mathsf{x}_{i}$ has an intended receiver located at fixed distance $d$. The fixed distance assumption, which can be seen as a \emph{target} transmission distance dictated by the network protocol, is commonly accepted, see for example \cite{weber10}. In order to measure the (spatially-averaged) link performance in the network we define a \emph{reference link}, cf. \cite{weber05}; the reference link consists of a reference receiver placed at an \emph{arbitrary} location $y_{0}\in\mathbb{R}^2$ and of an associated reference transmitter placed at $x_{0}$, where $x_{0}$ lies somewhere on the circle $\partial b(y_{0},d)$. Note that neither the receiver $y_{0}$ nor the transmitter $x_{0}$ are part of the point process $\Phi$. By the Slivnyak-Mecke Theorem \cite{stoyan95}, the statistics of $\Phi$ are not affected by the addition of the reference link. % which is due to the Poisson nature of $\Phi$.  %When doing this conditioning, we have to ensure that the event $\mathsf{x}_{0}=x_{0}$ is inside the set of possible outcomes of $\Phi$. Since $\lambda(x)$ has support $\mathbb{R}^2$ this is indeed always the case, though the event $\mathsf{x}_{0}=x_{0}$ may have zero probability, i.e., whenever $\lambda(x_{0})=0$. 
%Note that the reference link $x_{0}\to y_{0}$ is typical in the sense that it reflects the typical link performance for transmissions over distance $d$ with receivers located at distance $|y_{0}|$ to the origin.

We consider a path loss plus block-fading channel with independent and identically distributed (i.i.d.) fading coefficients. The power path loss between two locations $x,y\in\mathbb{R}^2$ is given by the path loss function $\ell(\|x-y\|)\triangleq(c+\|x-y\|^{\alpha})^{-1}$ with path loss exponent $\alpha\geq 2$, where $c\geq0$ ensures boundedness of $\ell$. The power fading coefficient between a transmitter at $x$ and the reference receiver $y_{0}$ is given by $\mathsf{g}_{x}$, where $\mathbb{E}\left[\mathsf{g}_{x}\right]=1$ for all $x$. When appropriate we will drop the index $x$ in $\mathsf{g}_{x}$.

\begin{figure}[t]
	\psfrag{a}[c][c][1][-50]{\small{$F(r)$}}
	\psfrag{b}[c][c]{\small{$y_{0}$}}
	\psfrag{c}[c][c]{\small{$x_{0}$}}
	\psfrag{d}[c][c]{\small{$d$}}
	  \centering
    \includegraphics[width=0.49\textwidth]{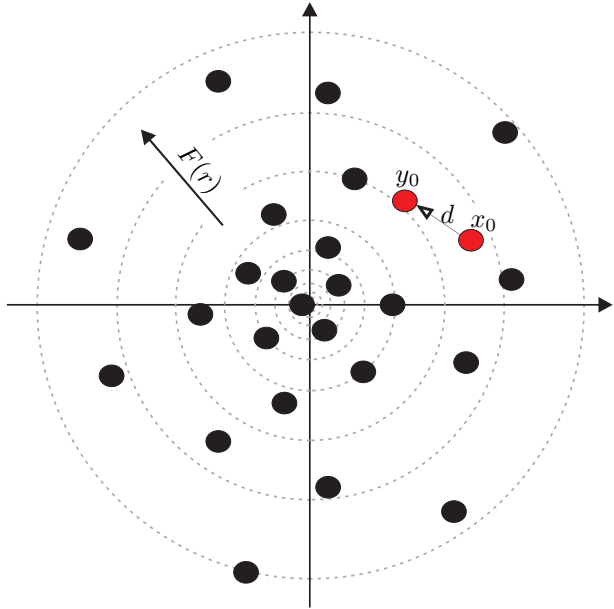}
\caption{System model: The reference receiver is located at $y_{0}$ with distance $\|y_{0}\|$ to the origin. The associated reference transmitter is located $d$ units away from $y_{0}$ at location $x_{0}$. Black dots represent interferers with random locations $\mathsf{x}_{1},\mathsf{x}_{2},\ldots$. Shape function $F(r)$ characterizes density of interferer set $\Phi$ over distance $r$ to the origin.}
\label{fig:typical_link}
\end{figure}

We assume that all nodes transmit with the same fixed transmit power and at a common information rate. The sum interference power at the reference receiver $y_{0}$ is then given by
\begin{IEEEeqnarray}{c}
	\mathsf{I}(y_{0})\triangleq\sum\limits_{ \mathclap{\mathsf{x}\in\Phi} }\mathsf{g}_{\mathsf{x}}\ell(\|\mathsf{x}-y_{0}\|).
\end{IEEEeqnarray}
%Intuitively, the reference transmitter $x_{0}$ is removed from $\Phi$ since this transmitter does not contribute to it. 
%It is further assumed that the sum interference \emph{signal} is conditionally additive white Gaussian noise (AWGN), i.e., it follows a zero-mean Gaussian distribution with random variance $\mathsf{I}(y_{0})$. In other words, conditioning on a spatial configuration of interferers $x_{1},x_{2},\ldots$ as well as on each transmitter's channel gain to the reference receiver $g_{x_{1}y_{0}},g_{x_{2}y_{0}},\ldots$, the sum interference signal is zero-mean Gaussian distributed with (deterministic) variance $I(y_{0})$.

Treating interference as white noise, the instantaneous signal-to-interference-plus-noise ratio $\sinr$ at the reference receiver $y_{0}$ is given by
\begin{IEEEeqnarray}{rCl}
	\sinr(y_{0})&\triangleq&\frac{\mathsf{g}_{x_{0}}}{\tfrac{1}{\eta}+\ell(d)^{-1}\mathsf{I}(y_{0})},\IEEEeqnarraynumspace\label{eq:sinr}
\end{IEEEeqnarray}
where $\eta$ is the average signal-to-noise ratio. We assume that the nodes employ strong channel coding such that the outage event is a steep function of the $\sinr$. Focusing on the case where explicit transmitter coordination as well as CSI feedback is not possible, the outage probability is a useful metric for characterizing the link performance.% In such a scenario the transmitter is blind to the instantaneous realization of both the interference power $\mathsf{I}(y_{0})$ and the channel fading $\mathsf{g}_{x_{0}}$ which may eventually lead to link \emph{outage}.

\begin{definition}
The outage probability of the reference link $x_{0}\to y_{0}$ is given by% the reduced Palm probability
\begin{IEEEeqnarray}{rCl}\label{eq:outage}
	\op(y_{0})\triangleq\mathbb{P}\left(\sinr(y_{0})<\beta\right), \label{eq:def_op}
\end{IEEEeqnarray}
where $\beta$ is a modulation and coding specific threshold. 
\end{definition}

\subsection{Spatial shapes chosen for illustrations}\label{sec:spatial_shapes}
\begin{figure}[t]
	\psfrag{tag1}[c][c]{\small{$r$}}
	\psfrag{tag2}[c][c]{\small{$F(r)$}}
	\psfrag{a}[c][c]{\small{a)}}
	\psfrag{b}[c][c]{\small{b)}}
	\psfrag{c}[c][c]{\small{c)}}
	\psfrag{d}[c][c]{\small{d)}}
		   \centering
    \includegraphics[width=0.5\textwidth]{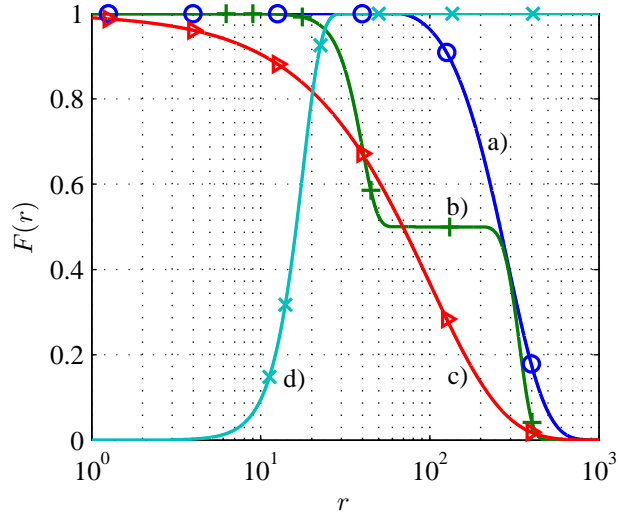}
\caption{The four exemplary spatial shapes considered in this work. Scenario a) (finite network), Scenario b) (urban with hotspot), Scenario c) (scattered decentralized network) and Scenario d) (carrier sensing in decentralized networks).}
\label{fig:scenarios}
\end{figure}

We next introduce four exemplary spatial shapes used for illustrations and numerical evaluations. They are chosen such to roughly characterize typical scenarios in wireless networks and to help increasing the reader's intuition about the results. An exact validation of the chosen spatial models through comparison with real-world deployments is outside the scope of this paper. The spatial shapes are depicted in Fig. \ref{fig:scenarios}.
\begin{itemize}
	\item \textit{Scenario a), finite network:} This scenario reflects the basic property all practical networks share: the network area is finite, or equivalently, the node density tends to zero for sufficiently large distances to the network center. It is assumed that the density first remains constant over a wide range as in the stationary case. At the network boundary the density then starts to decay rapidly until it becomes zero.
	\item \textit{Scenario b), urban with hotspot:} In urban scenarios there may sometimes exist small areas with very high data traffic, i.e., communications hotspots. They are typically found in commercial areas or other public places, comprising many densely--and sometimes dynamically--deployed wireless architectures \cite{andrews12}. We model such a hotspot scenario by ``adding a plateau of density'' around the origin on top of an already existing level of density corresponding to the urban deployment. This level then decays to zero with increasing distance to the origin to reflect finiteness of the network.    
	\item \textit{Scenario c), scattered decentralized network:} There are certain types of applications that preclude a detailed network layout for the reason of hostile environments or limited geographic access. For instance, large sensor networks are sometimes created by airdrop which results in a highly scattered spatial distribution of devices. We model such a behavior by an exponentially decreasing density around the origin.
	\item \textit{Scenario d), carrier sensing in decentralized networks:} This scenario is found in decentralized networks with transmitters employing carrier sensing to avoid excessive interference; suppose a transmitter (here located in the origin) is granted access to the medium. Consequently, other potential transmitters directly surrounding this transmitter defer their transmission as they sense the medium as busy, while other potential transmitters farther away sense the medium as free and therefore start to transmit, cf. Section \ref{sec:applications_inhibition} for more details.% As a result, the density of active transmitters around the considered transmitter at $o$ behaves effectively as indicated by Scenario d) in Fig. \ref{fig:scenarios}.
\end{itemize}

\section{Interference Analysis}\label{sec:interference}

%\newcounter{tmpeqnnum}
%\begin{figure*}[!b]
%\normalsize
%\vspace*{4pt}
%\hrulefill
%\setcounter{tmpeqnnum}{\value{equation}}
%\begin{IEEEeqnarray}{c}
%\setcounter{equation}{12}
%A_{4}(y_{0},c)=\frac{\pi}{2\sqrt{c}}\left(F(r)\,\text{arctan}\frac{2\mathfrak{R}\{\kappa(r,c,y_{0})\}}{1-|\kappa(r,c,y_{0})|^2}
%\bigg\vert_{r=0}^{\infty}-\int_{0}^{\infty} f(r)\,\text{arctan}\frac{2\mathfrak{R}\{\kappa(r,c,y_{0})\}}{1-|\kappa(r,c,y_{0})|^2}\,\mathrm dr\right)\label{eq:A4}
%\end{IEEEeqnarray}
%\setcounter{equation}{\value{tmpeqnnum}}
%\end{figure*}
We now study the interference statistics at the reference receiver $y_{0}$. The analysis first focuses on the case of an arbitrary fading distribution. We derive the first moment of the interference and then use bounding techniques such as those used in \cite{weber10} to characterize the interference distribution. For the Rayleigh model, we then derive the Laplace transform of the interference distribution.

\subsection{Arbitrary Fading Model}\label{sec:int_general_fading}

\subsubsection{First Moment of the Interference}
The first moment of the interference $\mathbb{E}\left[\mathsf{I}(y_{0})\right]$ measured at the reference receiver $y_{0}$ can in general be obtained by
\begin{IEEEeqnarray}{rCl}
	\int_{0}^{\infty} \mathbb{P}\left(\mathsf{I}(y_{0})\geq z\right)\,\mathrm dz.\label{eq:first_moment_int}
\end{IEEEeqnarray}
Obviously, one would have to know the distribution of $\mathsf{I}(y_{0})$, which unfortunately is known in closed-form only for a few cases of stationary point processes \cite{ganti09}. A remedy to this problem is given by the Campbell Theorem \cite{stoyan95}, which allows us to derive the first result:   

\begin{theorem}\label{theorem:moment1}
Let $f(r)\triangleq\mathrm d F(r)/\mathrm d r$, %\footnote{In cases where $F(r)$ is not differentiable, e.g., $F(r)=1-H(r)$ where $H(r)$ is the Heaviside function \cite{olver10}, the differentiation is understood within the context of distribution theory.} 
$c>0$, $\alpha=2$ and $\|y_{0}\|>0$. If $ F(r)\widesim{r\to\infty}\frac{1}{r^\nu}$ for some $\nu>0$, then
	\begin{IEEEeqnarray}{c}
		\mathbb{E}\left[\mathsf{I}(y_{0})\right] = \lambda A_{2}(y_{0},c)<\infty,\label{eq:exp_int2a}
	\end{IEEEeqnarray}
	where $A_{2}(y_{0},c)$ is given by
	\begin{IEEEeqnarray}{c}
		A_{2}(y_{0},c)=-\pi \left(F(0)\,\text{\textnormal{asinh}}\frac{c-\|y_{0}\|^2}{2\|y_{0}\|\sqrt{c}}+\int_{0}^{\infty}f(r)\,\text{\textnormal{asinh}}\frac{r^2+c-\|y_{0}\|^2}{2\|y_{0}\|\sqrt{c}}\,\mathrm dr\right).\IEEEeqnarraynumspace\label{eq:A2}
	\end{IEEEeqnarray}
\end{theorem}
A proof is given in Appendix \ref{sec:proofs}. The condition $ F(r)\widesim{r\to\infty}\frac{1}{r^\nu}$ for some $\nu>0$ is necessary for $\mathbb{E}\left[\mathsf{I}(y_{0})\right]$ to exist. The function $A_{2}(y_{0},c)$ in (\ref{eq:A2}) has an interesting interpretation: it can be seen as the \emph{interference-driving} function as it determines the interference up to a scaling factor. Additionally, the first term in (\ref{eq:A2}) can be interpreted as the interference field associated with the origin $o$, while the second term effectively adds up the interference according to $f(r)$.

\begin{remark}\label{rem:a2}
	If the reference receiver is located in the origin ($\|y_{0}\|=0$), the $\text{\textnormal{asinh}}$-term in $A_{2}(y_{0},c)$ has to be replaced by $\log (r^2 + \|y_{0}\|^2+c)$, cf. Identity 2 in Appendix \ref{sec:integrals}.
\end{remark}
Glancing at the second term in (\ref{eq:A2}), we note:
\begin{corollary}\label{col:random_distance}
	When $F(0)=1$ and $f(r)\leq 0$ for all $r\in\mathbb{R}_{+}$, $F(r)$ can be interpreted as a complementary cumulative distribution function (CDF) with respect to a \emph{random} distance $\mathsf{r}$ to the origin, yielding
		\begin{IEEEeqnarray}{c}
			A_{2}(y_{0},c) = -\pi\,\text{\textnormal{asinh}}\frac{c-\|y_{0}\|^2}{2\|y_{0}\|\sqrt{c}}+\pi\mathbb{E}\left[\text{\textnormal{asinh}}\frac{\mathsf{r}^2+c-\|y_{0}\|^2}{2\|y_{0}\|\sqrt{c}}\right].\IEEEeqnarraynumspace
		\end{IEEEeqnarray}
\end{corollary}
Corollary \ref{col:random_distance} states that the integral in (\ref{eq:A2}) can be seen as an \emph{averaging} of the interference with respect to a random distance $\mathsf{r}$. Such a representation may be appropriate when analyzing networks with \emph{a priori} unknown or fast-varying spatial configurations, for which a CDF is then used to model their spatial shape.

% \begin{corollary}
% Letting $|y_{0}|\to 0$, we further have
% 		\begin{IEEEeqnarray}{c}
% 			A_{2}(0,c) = \log(1/2c)+\mathbb{E}\left[\log(2(\mathsf{r}+c))\right].\IEEEnonumber
% 		\end{IEEEeqnarray}
% \end{corollary}

\begin{corollary}\label{col:sparse_network}
	Let $ F(r)\widesim{r\to\infty}\frac{1}{r^\nu}$ for some $\nu\in(0,2]$. Then, the expected number of interferers $\mathbb{E}\left[|\Phi|\right]=2\pi\lambda\int_{0}^{\infty}rF(r)\,\mathrm dr=\infty$ but the expected interference $\mathbb{E}\left[\mathsf{I}(y_{0})\right]<\infty$.
\end{corollary}

The intuition behind Corollary \ref{col:sparse_network} is that, although the expected number of nodes contributing to the interference is unbounded, the network remains sufficiently sparse such that the first moment of the interference remains bounded. Note that for a PPP, if the expected number of interferers is unbounded, this implies that the number of interferers is a.s. infinite which can be verified by studying the Laplace transform of the PPP \cite{stoyan95,haenggi12}. Applying the Markov Inequality %\footnote{which shows that $\mathbb{E}\left[\mathsf{I}(y_{0})\right]<\infty$ implies that a.s. $\mathsf{I}(y_{0})<\infty$.} 
$\mathbb{P}(\mathsf{z}\geq z)\leq \tfrac{1}{z}\mathbb{E}\left[\mathsf{z}\right]$ \cite{gut05}, this in turn means that the number of nodes is a.s. infinite while the interference remains a.s. finite. This particular finding is somewhat remarkable since it rearranges the commonly-accepted perception, stating that whenever $\alpha=2$ and the number of interferers is a.s. infinite, the interference is a.s. infinite as well \cite{ganti09, weber05}. This perception indeed holds for stationary point processes but does not hold in general for non-stationary point processes, as demonstrated by Corollary \ref{col:sparse_network}.

\begin{theorem}\label{thm:dense_network}
	Let $ F(r)\widesim{r\to\infty}\frac{1}{r^\nu}$ for $\nu\to0$. Then, $\mathsf{I}(y_{0})=\infty$ a.s.
\end{theorem}
A proof is given in Appendix \ref{sec:proofs}. Theorem \ref{thm:dense_network} shows that whenever $F(r)$ decays at most logarithmically, the interference is a.s. infinite. In particular, this includes the stationary case (since $\lim_{r\to\infty}F(r)>0$) which is consistent with the literature \cite{ganti09}. Combining Corollary \ref{col:sparse_network} and Theorem \ref{thm:dense_network}, we observe that for asymptotically decaying $F(r)$ there exists a transition between \emph{sparse} and \emph{dense} networks. This transition determines whether or not the interference is a.s. finite in a non-stationary Poisson network with a.s. infinite number of interferers.

%\begin{remark}
%Note that by Markov's inequality $\mathbb{P}(\mathsf{z}\geq z)\leq \tfrac{1}{z}\mathbb{E}\left[\mathsf{z}\right]$, so $\mathbb{E}\left[\mathsf{I}(y_{0})\right]<\infty$ implies that $\mathsf{I}(y_{0})<\infty$ a.s. Hence, the above results can be transformed into a.s. convergence results.
%\end{remark}

\begin{remark}
By setting $\mathsf{g}\equiv1$, the pure path loss case is also covered by the above results. 
\end{remark}

We now characterize the first moment of the interference for the case $\alpha=4$.
\begin{theorem}\label{theorem:moment2}
Let $f(r)\triangleq\mathrm d F(r)/\mathrm d r$, $c>0$ and $\alpha=4$. Then,
\begin{IEEEeqnarray}{c}
	\mathbb{E}\left[\mathsf{I}(y_{0})\right] = \lambda\,A_{4}(y_{0},c)<\infty,\IEEEeqnarraynumspace\label{eq:exp_int2}
\end{IEEEeqnarray}
	where $A_{4}(y_{0},c)$ is given by
	\begin{IEEEeqnarray}{c}
		A_{4}(y_{0},c)=\frac{\pi}{2\sqrt{c}}\left(F(r)\,\text{\textnormal{arctan}}\frac{2\mathfrak{R}\{\kappa(r,c,y_{0})\}}{1-|\kappa(r,c,y_{0})|^2}
		\Big\vert_{r=0}^{\infty}-\int_{0}^{\infty} f(r)\,\text{\textnormal{arctan}}\frac{2\mathfrak{R}\{\kappa(r,c,y_{0})\}}{1-|\kappa(r,c,y_{0})|^2}\,\mathrm dr\right)\label{eq:A4}\IEEEeqnarraynumspace
	\end{IEEEeqnarray}
	and $\kappa(r,c,y_{0})$ is given by (\ref{eq:kappa}) in Appendix \ref{sec:integrals}.
\end{theorem}
A proof is given in Appendix \ref{sec:proofs}. The term $A_{4}(y_{0},c)$ in (\ref{eq:A4}) can be again interpreted as the interference-driving function.

\begin{corollary}\label{col:homogeneous}
	Let $F(r)=1$ so that $f(r)=0$ (stationary PPP). Then, by carefully taking the limits
		\begin{IEEEeqnarray}{c}
			\lim\limits_{r\to a}\arctan\frac{2\mathfrak{R}\{\kappa(r,c,y_{0})\}}{1-|\kappa(r,c,y_{0})|^2}=
					\begin{cases}
						-\frac{\pi}{2}, & a=0\\
                   \frac{\pi}{2}, & a=\infty,
					\end{cases}\IEEEnonumber
		\end{IEEEeqnarray}
		the well-known result $\mathbb{E}\left[\mathsf{I}(y_{0})\right]=\lambda\frac{\pi^2}{2\sqrt{c}}$ for the stationary PPP is recovered \cite{ganti09}.
\end{corollary}

\subsubsection{Bounds on the Interference Distribution}
We next treat the problem of bounding the tail of $\mathbb{P}\left(\mathsf{I}(y_{0})\geq z\right)$. A simple upper bound can be obtained using Markov's inequality in combination with Theorem \ref{theorem:moment1} and Theorem \ref{theorem:moment2}. For the construction of a lower bound, we first recall the definition of subharmonic functions.

\begin{definition}{(Subharmonic functions \cite[Ch. 2]{krantz92}):}
	Let $G\subseteq\mathbb{R}^n$ be an open set and let $h(x)$ be a function twice continuously differentiable on $G$. If $\sum_{k=1}^{n}\frac{\partial^2}{\partial x_{k}^2}h(x)\geq0$ on $G$, then $h(x)$ is called subharmonic on $G$.
\end{definition}

If $F(r)$ is convex in a certain (one-dimensional) region, then the intensity $\lambda(x)$ is subharmonic on the corresponding (two-dimensional) region. Such a behavior may be often found at the network boundary, e.g., Scenario a) and b), or when the shape function exhibits a tail, e.g., Scenario c) and d). In this case a lower bound on the tail probability $\mathbb{P}\left(\mathsf{I}(y_{0})\geq z\right)$ can be derived:

\begin{theorem}\label{theorem:lower_bound}
	Let $\lambda(x)$ be subharmonic on $G\subseteq\mathbb{R}^2$ and let $y_{0}\in G$. Denote by $\bar r(x)$ the maximum radius for which the closed ball $b(x,t)$ is contained in $G$, i.e., $\bar r(x)=\max\{t\in\mathbb{R}_{+}:b(x,t)\subseteq G\}$. Then,% $\mathbb{P}\left(\mathsf{I}(y_{0})\geq z\right)$ can be bounded as
	\begin{IEEEeqnarray}{c}
		\mathbb{P}\left(\mathsf{I}(y_{0})\geq z\right)\geq1-\exp\left(-2\pi\lambda F(\|y_{0}\|)\int_{0}^{\bar r(y_{0})}r\,\mathbb{P}\left(\mathsf{g}\geq z (c+r^{\alpha})\right)\,\mathrm dr\right).\IEEEeqnarraynumspace\label{eq:lb_subharmonic_general}
	\end{IEEEeqnarray}
\end{theorem}
A proof is given in Appendix \ref{sec:proofs}. Note that subharmonicity includes the case of harmononicity. The construction of the lower bound in Theorem \ref{theorem:lower_bound} basically builds on the so-called ``dominant interferer'' phenomenon introduced in \cite{weber05}, where it was also reported that the resulting bound is fairly tight. However, in our case the tightness of (\ref{eq:lb_subharmonic_general}) strongly depends on the second derivative of $F(r)$ and may not be guaranteed.

\begin{corollary}
	Let $z<\tfrac{1}{c}$. For the pure path loss model ($\mathsf{g}\equiv1$), (\ref{eq:lb_subharmonic_general}) reduces to
	\begin{IEEEeqnarray}{c}
		\mathbb{P}\left(\mathsf{I}(y_{0})\geq z\right)\geq1-\exp\left(-\pi\lambda F(\|y_{0}\|)\min\big\{\bar r^2(y_{0}), (\tfrac{1}{z}-c)^{\frac{2}{\alpha}}\big\}\right).\IEEEeqnarraynumspace\label{eq:lower_bound_inter_col}
	\end{IEEEeqnarray}
\end{corollary}
The restriction $z<\tfrac{1}{c}$ is necessary to allow the closest interferer to be dominant, otherwise this bounding technique would yield the trivial lower bound $\mathbb{P}\left(\mathsf{I}(y_{0})\geq z\right)\geq0$. %Consequently, the bound in (\ref{eq:lower_bound_inter_col}) is useful either for low $z$, e.g., high target SINRs $\beta$, or when setting $c=0$ (singular path loss model).

Using the Markov Inequality \cite{gut05}, we obtain the simple upper bound
\begin{IEEEeqnarray}{c}
	\mathbb{P}\left(\mathsf{I}(y_{0})\geq z\right) \leq \frac{\lambda}{z} A_{\alpha}(y_{0},c),
\end{IEEEeqnarray}
for the cases $\alpha=2$ and $\alpha=4$, where $A_{2}(y_{0},c)$ and $A_{4}(y_{0},c)$ are given by (\ref{eq:A2}) and (\ref{eq:A4}), respectively.

\subsection{Rayleigh Fading Model}\label{sec:int_rayleigh_fading}
For the commonly-used Rayleigh fading model, it was demonstrated in \cite{baccelli06} that the Laplace transform of the interference is useful for computing outage probabilities. % can be obtained in closed-form via the Laplace transform of the interference.\footnote{Theoretically, it suffices that only the reference transmission $x_{0}\to y_{0}$ is subject to Rayleigh fading to exploit the Laplace transform of $\mathsf{I}(y_{0})$.} 
We therefore derive the Laplace transform of $\mathsf{I}(y_{0})$, i.e., $\mathcal{L}_{\mathsf{I}(y_{0})}(s)=\mathbb{E}\left[e^{-s\mathsf{I}(y_{0})}\right]$ next for arbitrary $y_{0}$ and $F(r)$. Similar to Section~\ref{sec:int_general_fading}, we focus again on $\alpha=2$ and $\alpha=4$.

\begin{theorem}\label{theorem:laplace_int}
	In the Rayleigh fading model, the Laplace transform of $\mathsf{I}(y_{0})$ at $y_{0}$ is given by
	\begin{IEEEeqnarray}{c}
		\mathcal{L}_{\mathsf{I}(y_{0})}(s)=\exp\left(-\lambda s \,A_{\alpha}(y_{0},s+c)\right),\IEEEeqnarraynumspace\label{eq:laplace}
	\end{IEEEeqnarray}
	for the cases $\alpha=2$ and $\alpha=4$, where $A_{2}(y_{0},c)$ and $A_{4}(y_{0},c)$ are given by (\ref{eq:A2}) and (\ref{eq:A4}), respectively.
\end{theorem}

A proof is given in Appendix \ref{sec:proofs}. Note that for $\alpha=2$ and $ F(r)\widesim{r\to\infty}\frac{1}{r^\nu}$ for every $\nu\to0$, we have that $\mathcal{L}_{\mathsf{I}(y_{0})}(s)=0$ for all $s$. This in turn implies $\mathsf{I}(y_{0})=\infty$ a.s. which is consistent with Theorem \ref{thm:dense_network}.

\begin{remark}
Setting $F(r)=1$ for all $r\in\mathbb{R}_{+}$ and $c=0$, we recover the well-known result for the homogeneous case with $\alpha=4$: $\mathcal{L}_{\mathsf{I}(y_{0})}(s)=\exp(-\lambda\frac{\pi^2}{2} \sqrt{s})$.
\end{remark}

\begin{remark}
 The case $\alpha=2$ with Rayleigh fading may seem contradictory first: A path loss exponent equal to $\alpha=2$ is typically observed in propagation environments without ground-plane reflection \cite{tse05}. In contrast, Rayleigh fading models the non-line-of-sight (NLOS) case with many reflected paths impinging at the receiver. It turns out, however, that there may truly exist urban NLOS scenarios with considerably small path loss exponents ($\alpha\sim2.6$) as demonstrated in \cite{feuerstein94}. Depending on the geometry of objects in the proximity of the receiver, the received signal may therefore still undergo severe small-scale fading. Hence, the case $\alpha=2$ may serve as a theoretical limit of what can be expected roughly in Rayleigh fading environments with a small path loss exponent.
\end{remark}

\section{Outage Probability and Model Accuracy}\label{sec:op_accuracy}
In this section, the outage probability is characterized for the underlying setting and the model accuracy is studied.%Outage probability as defined in (\ref{eq:outage}) is an important physical layer metric for characterizing the performance of the network; clearly, if the outage probability is undesirably high the whole communication process will be degraded irrespective of the amount of engineering effort on the upper layers.% For this reason outage probability is often considered as a bottleneck metric. It is thus worth to study the outage probability in detail. 

\subsection{Outage Probability}\label{sec:op}
For arbitrary fading the bounds on the tail probability of $\mathsf{I}(y_{0})$ can be used to bound the outage probability, see for example \cite{weber05}. In the sequel, we will focus on the Rayleigh fading case and discuss the impact of the spatial shape on the resulting performance.% As already mentioned, the Laplace transform of the interference power can be used to compute the outage probability. We therefore note:

\begin{corollary}\label{col:op_rayleigh}
The outage probability $\op(y_{0})$ at $y_{0}$ in the Rayleigh fading model is
	\begin{IEEEeqnarray}{c}
		\op(y_{0})=1-\mathcal{L}_{\mathsf{I}(y_{0})}\left(\beta\,(c+d^{\alpha})\right)e^{-\frac{\beta}{\eta}},\IEEEeqnarraynumspace\label{eq:op}
	\end{IEEEeqnarray}
	for the cases $\alpha=2$ and $\alpha=4$, where $\mathcal{L}_{\mathsf{I}(y_{0})}$ is given by (\ref{eq:laplace}).\label{col:ray_out}
\end{corollary}

\begin{figure}[t]
	\psfrag{tag1}[c][c]{\small{$\eta=\infty$}}
	\psfrag{tag2}[c][c]{\small{$\eta=10$ dB}}
	\psfrag{tag3}[c][c]{\small{$\|y_{0}\|$}}
	\psfrag{tag4}[c][c]{\small{$\op(y_{0})$}}
	\psfrag{tag5ta}{\footnotesize{$\alpha=2$}}
	\psfrag{tag6}{\footnotesize{$\alpha=4$}}
		   \centering
    \includegraphics[width=0.5\textwidth]{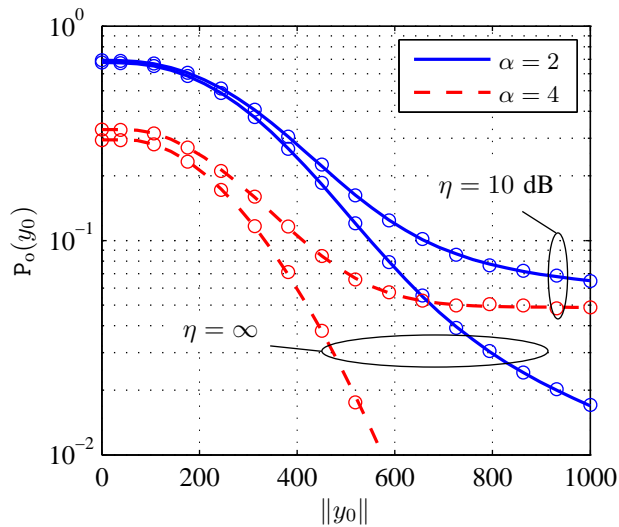}
\caption{Outage probability $\op(y_{0})$ vs. $\|y_{0}\|$. Parameters were $d=10$, $\beta=0.5$, $c=1$, $\lambda=0.001$. The spatial shape function $F(r)$ was chosen according to scenario a). Marks represent simulation results.}
\label{fig:op}
\end{figure}

%We present a proof in Appendix \ref{sec:proofs} for the reason of rigorousness. 
%The technique for obtaining Corollary \ref{col:op_rayleigh} is well-known in the literature, cf. \cite{weber05,ganti09}. It builds on the Laplace transform $\mathcal{L}_{\mathsf{I}(y_{0})}(s)$ and on exploiting the fact that the $\mathsf{g}$ are exponentially distributed. 

Corollary~\ref{col:ray_out} now allows for measuring the exact outage probability at an arbitrary location $y_{0}$ and for an arbitrary spatial shape function $F(r)$. Fig.~\ref{fig:op} shows $\op(y_{0})$ vs. $\|y_{0}\|$ for $\alpha=2$ and $\alpha=4$. The spatial shape function $F(r)$ was chosen according to scenario a). In the noise-free case ($\eta=\infty$) the outage probability decreases monotonically with increasing distance to the origin. Furthermore, we observe that for $\alpha=2$ the outage probability $\op(y_{0})$ is higher and its slope is less steep than it is for $\alpha=4$. This is because for $\alpha=2$ the individual interference contributions decay more slowly over distance than they do for $\alpha=4$. As a result, the interference is no longer dominated by only a few nearby interferers but it is determined by the large number of nodes nodes, including those relatively far away from $y_{0}$. When receiver noise is considered ($\eta=10$ dB) the behavior of $\op(y_{0})$ changes considerably: while $\op(y_{0})$ is on the same order as in the noise-free case around the center of the network, both curves converge to a constant outage probability level as the reference receiver eaves the center of the network. In fact, in this boundary region outage is primarily due to bad fades rather than to interference, thus rendering the performance noise-limited rather than interference-limited. This transition--from the interference-limited to the noise-limited regime--can be precisely tracked owing to the developed model; for instance, Fig.~\ref{fig:scenarios} suggests that the noise-limited regime commences somewhere around $\|y_{0}\|\approx500$, while Fig.~\ref{fig:op} reveals that this is not true at least for $\alpha=2$ ($\|y_{0}\|>800)$ for the reason explained above.

\subsection{Exact vs. Approximate Model}\label{sec:accuracy}

Up to this point, it is not yet very clear how much can be gained by the interference model derived in this work. In order to quantify the gains, we compare our model to a simpler one that approximates the non-stationarity-property of the interference locally around the reference receiver $y_{0}$. In this simpler model, the interference field at $y_{0}$ is assumed to originate from a stationary PPP having constant intensity $\lambda F(\|y_{0}\|)$; in other words, the network-wide spatial distribution of interferers is approximated locally by the density at location $y_{0}$.   

\begin{figure*}[!t]
	\psfrag{tag1}[c][c]{\small{$\|y_{0}\|$}}
	\psfrag{tag2}[c][c]{\small{$\gamma(y_0)$}}
	\psfrag{tagerror}[c][c]{\small{$\delta(y_{0})$}}
	\psfrag{a}[c][c]{\small{a)}}
	\psfrag{b}[c][c]{\small{b)}}
	\psfrag{c}[c][c]{\small{c)}}
	\psfrag{d}[c][c]{\small{d)}}
	\centerline{\subfloat[Log-divergence]{\includegraphics[width=0.5\textwidth]{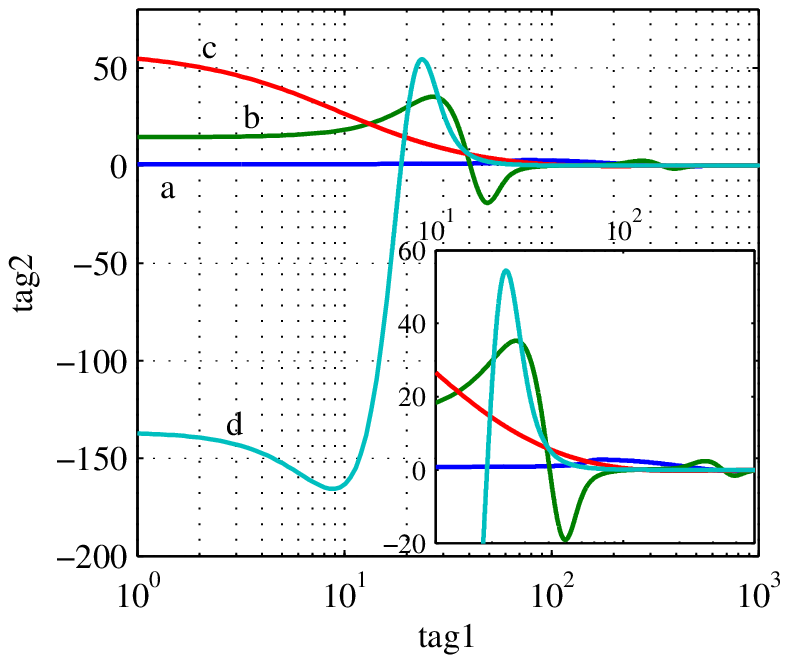}
	\label{fig:gamma}}
	\hfil
	\subfloat[Relative approximation error]{\includegraphics[width=0.5\textwidth]{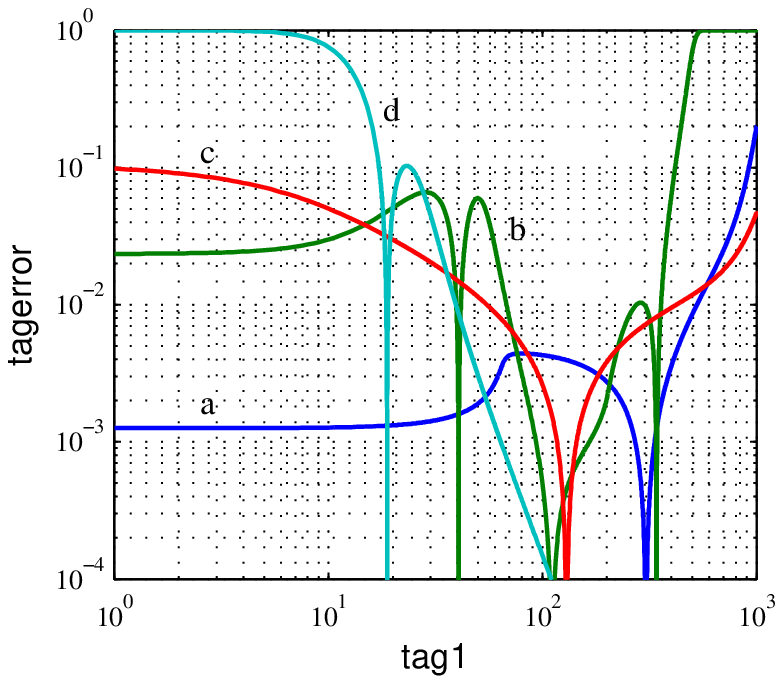}
	\label{fig:rel_error}}}
	\caption{(a) Log-divergence $\gamma(y_{0})$ vs. $\|y_{0}\|$ for different shape functions. (b) Relative approximation error vs. $\|y_{0}\|$ for different shape functions and $\lambda=10^{-3}$. Parameters are $\alpha=4$, $c=0$, $ d=10$ and $\beta=1$.}
\end{figure*}

For the stationary case the outage probability for Rayleigh fading is well-known \cite{baccelli03,baccelli06}. With the above approximation the approximate outage probability $\opt(y_{0})$ at $y_{0}$ is then given by
\begin{IEEEeqnarray}{c}
\opt(y_{0})=1-\exp\left(-\lambda F(\|y_{0}\|) d^2\beta^{\frac{2}{\alpha}}\tfrac{2\pi^2}{\alpha}\csc\tfrac{2\pi}{\alpha}\right).\IEEEeqnarraynumspace\label{eq:approx_rayleigh_op}
\end{IEEEeqnarray}
The intuition behind (\ref{eq:approx_rayleigh_op}) is that the exact outage probability is approximated by taking the outage probability expression corresponding to the stationary case and modulating the intensity $\lambda$ by the spatial shape function $F(r)$ at $r=\|y_{0}\|$. To measure the difference between $\opt(y_{0})$ and $\op(y_{0})$ we define the following metric.
\begin{definition}[Log-divergence]
The log-divergence is defined as
\begin{IEEEeqnarray}{c}
	\gamma(y_{0})\triangleq\lambda^{-1}\log\frac{1-\op(y_{0})}{1-\opt(y_{0})}.
\end{IEEEeqnarray}
\end{definition}
The log-divergence $\gamma(y_{0})$ quantifies the ratio of the exact and approximate success probability ($1-\op(y_{0})$) on the logarithmic scale for arbitrary $y_{0}$. The normalization by $\lambda$ is necessary to remove the dependency on $\lambda$ so to measure the divergence resulting from the spatial shape only. % Since the actual divergence technically scales with $\lambda$, $\gamma(y_{0})$ should be interpreted as the logarithmic success probability divergence per density $\lambda$. 
For large positive $\gamma(y_{0})$ the approximation overestimates the true outage probability, while for large negative $\gamma(y_{0})$ outage probability is underestimated. The approximation is accurate whenever $|\gamma(y_{0})|$ is small. The log-divergence can be computed using Corollary~\ref{col:op_rayleigh}.
\begin{corollary}
	In the Rayleigh fading model and for $c=0$, the log-divergence for the case $\alpha=4$ is
	\begin{IEEEeqnarray}{c}
		\gamma(y_{0})=d^{4}\beta\left(\frac{\pi^2F(\|y_{0}\|)}{2d^{2}\sqrt{\beta}}-A_{4}(y_{0},\beta d^{4})\right).\IEEEeqnarraynumspace
	\end{IEEEeqnarray}
\end{corollary}
Fig. \ref{fig:gamma} shows the log-divergence $\gamma(y_{0})$ vs. $\|y_{0}\|$ for the shape functions $F(r)$ introduced in Section \ref{sec:spatial_shapes}. It can be seen that the log-divergence exhibits an oscillatory behavior that depends on the degree of variability of $F(r)$. For example, $F(r)$ in scenario a) and c) changes comparably slow, and thus the corresponding log-divergence shows only weak oscillations. The log-divergence for scenario a) remains low ($\gamma(y_{0})\approx 0.80$) over a wide range which suggests that the local approximation works well in this case. As for scenario c), however, the log-divergence is large around the origin which is due to the fact that the outage probability is highly overestimated as the (exponential) decay of $F(r)$ to the right-hand side is neglected by the approximation. A similar effect can be observed for scenario d): the outage probability is highly underestimated around the origin because the increasing density to the right-hand side is neglected. The log-divergence for scenario b) exhibits rich oscillations due to a stronger varying shape function. As can be seen, these oscillations are high particularly in the transition region ($10^1\leq r\leq10^2$).% The latter also holds for scenario a) since the log-divergence increases by approximately 3.8 times around the transition point $r=10^2$ in this case.

To understand the impact of the log-divergence on outage probability, the relative error
\begin{IEEEeqnarray}{c}
	\delta(y_{0})=\frac{|\opt(y_{0})-\op(y_{0})|}{\op(y_{0})}\label{eq:rel_ap_error}
\end{IEEEeqnarray}
is shown in Fig. \ref{fig:rel_error} for the example $\lambda=10^{-3}$. %The relative error is an important measure for quantifying the outage probability deviation, and thus for characterizing the expected QoS variability. 
Fig. \ref{fig:rel_error} underlines the observations made in Fig. \ref{fig:gamma}: the approximation works well for scenario a) while for the other three scenarios the approximation is relatively loose. Especially for the scenarios b) and d), the relative error $\delta(y_{0})$ is considerably high over a wide range (between 1\% and 10\% around the network center for scenario c)). For scenario b) the relative error is approximately 10\% around the density mid-level. In case of scenario d) the approximation completely fails around the origin. %Interestingly, for the first three scenarios the relative error starts to increase rapidly when passing the transition point $r=10^2$. In contrast, the relative error decays for scenario d) as $\|y_{0}\|$ becomes large which is a result of $F(r)$ being asymptotically constant, cf. Fig. \ref{fig:scenarios}. Note that the zeros in Fig. \ref{fig:rel_error} correspond to the zeros in Fig. \ref{fig:gamma}.

\section{Applications and Examples}\label{sec:local_throughput}
In this section, the developed model is applied to problems in network modeling with non-stationary spatial node distributions.
\subsection{Local Transmission Capacity}
As argued in Section \ref{sec:introduction}, the transmission capacity metric cannot be applied to networks with non-stationary spatial node distributions. Based on the developed model, the definition of the transmission capacity can however be extended to account for non-stationarity and location dependency.
\begin{definition}[Local transmission capacity]
	The local transmission capacity is defined as
	\begin{IEEEeqnarray}{c}
		c(x,\epsilon)\triangleq\lambda(x,\epsilon)(1-\epsilon),\label{eq:def_dtc}
	\end{IEEEeqnarray}
	and gives the maximal density $\lambda(x,\epsilon)\triangleq\op^{-1}(x)$ of concurrent transmissions in an infinitesimal region around location $x$ subject to an outage probability constraint $\epsilon$.
\end{definition}
Since the local transmission capacity accounts for the spatial shape, this metric allows throughput bottlenecks to be spatially tracked and properly engineered, e.g., by balancing QoS among nodes irrespective of their location. For isotropic node distributions, in particular, the local transmission capacity depends on $\|x\|$ and can be computed/bounded by algebraic manipulations of the outage probability expressions derived in the previous sections.  %For Rayleigh fading, the local TC $c(x,\epsilon)$ is obtained by solving (\ref{eq:op}) for $\lambda$. Otherwise, the upper and lower bounds on the interference distribution can be used to obtain bounds on the local TC. %Note that the perhaps most desirable feature of the TC, which is the ability to benchmark different transmission protocols, is preserved by the local TC.

\textit{Example: FH-CDMA vs. DS-CDMA in decentralized networks:} In \cite{weber05} it was shown that in the stationary PPP model, the transmission capacity gain of frequency-hopping (FH)-CDMA compared to direct-sequence (DS)-CDMA scales as $M^{1-\frac{2}{\alpha}}$, where $M$ is the processing gain. For very small path loss exponents ($\alpha\to2$) this result suggests that the gains of FH-CDMA vanish irrespective of the processing gain $M$. This observation, however, results from the stationarity assumption not being able to correctly capture the case $\alpha=2$. The next result rearranges this scaling result for the case $\alpha=2$ for Rayleigh fading and the reference receiver located in the origin.
\begin{corollary}\label{col:fh_ds_gain}
	Let $\alpha=2$, $c=0$ and $\eta=\infty$. %\footnote{The requirements $c=0$ and $\eta=\infty$ are not imperative for the derivation of this Corollary and were chosen for the reason of comparison with \cite{weber05}.} 
	In the Rayleigh fading model, the local transmission capacity gain of FH-CDMA over DS-CDMA at $o$ is
	\begin{IEEEeqnarray}{c}
		\frac{c^{\text{FH}}(o,\epsilon)}{c^{\text{DS}}(o,\epsilon)}=1+\frac{F(0)}{A_{2}(o,\beta d^2)}\log M+O(1).
	\end{IEEEeqnarray}
\end{corollary}

A proof is given in Appendix \ref{sec:proofs}. Surprisingly, this result shows that the gains of FH-CDMA over DS-CDMA do not vanish for $\alpha=2$ (as was predicted by \cite{weber05}), but they scale with $\log M$ which re-enforces the superiority of FH-CDMA over DS-CDMA in terms of transmission capacity. %Although Corollary \ref{col:fh_ds_gain} assumes the reference receiver to be located in the origin, we conjecture that this scaling behavior is not changed when the reference receiver is located at an arbitrary location. Fig. \ref{fig:gain_fh_ds} supports our conjecture.

\subsection{Interference in Networks with Transmitter-Inhibition}\label{sec:applications_inhibition}
%Besides slotted Aloha, other MAC protocols, such as CSMA or local FDMA, represent effective techniques for reducing excessive interference generated by nearby nodes. These techniques coordinate transmissions locally to inhibit nodes from accessing resources that are already in use. 
In order to study decentralized networks with inhibition mechanisms such as carrier-sense medium access (CSMA) or local frequency division multiple access (FDMA) while ensuring analytic tractability, methods based on non-homogeneous Poisson approximation have been proposed for stationary models \cite{hunter10,baccelli09b,tanbourgi12,tanbourgi11_2}. When such protocols are \emph{transmitter-initiated}, e.g., transmitter sensing for CSMA, the resulting spatial distribution of interferers becomes non-homogeneous but remains isotropic around the inhibiting transmitter since potential transmitters around the inhibiting transmitter are kept silent while others located farther away are likely to transmit. In contrast, the interference field at the associated receiver is not isotropic. To overcome this intractability, the receiver is assumed to be co-located with the inhibiting transmitter, thereby virtually rendering the interference field around the receiver isotropic at the cost of losing model accuracy. This loss depends on the distance between the inhibiting transmitter and the associated receiver. Using the developed model, we can now evaluate the accuracy loss resulting from assuming that transmitter and receiver are co-located for this CSMA modeling technique. We briefly summarize the basic ideas of this modeling technique and refer to \cite{hunter10,tanbourgi12} for further details.

% \begin{figure}[t]
% 	\psfrag{a}[c][c]{\small{Active interferer}}
% 	\psfrag{b}[c][c]{\small{Inhibited interferer}}
% 	\psfrag{c}[c][c]{\small{$x_{0}$}}
% 	\psfrag{d}[c][c]{\small{$y_{0}$}}
% 	\centering
% 	\includegraphics[width=0.5\textwidth]{../figures/csma_motivation.eps}
% \caption{Snapshot: resulting interferer set after CSMA mechanism with transmitter-sensing. Grey dots represent inhibited interferers and black dots represent active interferers. An interferer is inhibited whenever it senses an ongoing transmission, here transmission of $x_{0}$. Interference field around receiver $y_{0}$ is not isotropic.}
% \label{fig:csma_motivation}
% \end{figure}

\textit{Non-homogenous Poisson approximation for CSMA networks:} Let the potential interferers be initially distributed according to a stationary PPP of density $\lambda$. Assume the that the reference transmitter $x_{0}$ and the reference receiver are located at $o$ and $y_{0}$, respectively, and separated by $d$. The inhibition mechanism is modeled in three steps: 1) The large-scale density of active interferers is derived using a Mat\'{e}rn-type 2 model \cite{stoyan95}, which captures the inhibition effect on a ``macroscopic'' level. The large-scale density $\lambda_{\ell}$ is then given by
\begin{IEEEeqnarray}{c}
	\lambda_{\ell}=\frac{1-e^{-\lambda\pi\Gamma(1+\tfrac{2}{\alpha})\Delta^{-\frac{2}{\alpha}}}}{\pi\Gamma(1+\tfrac{2}{\alpha})\Delta^{-\frac{2}{\alpha}}},
\end{IEEEeqnarray}
where $\Delta>0$ is the sensing threshold. We then condition on the fact that $x_{0}$ is granted access to the channel and we are now seeking the statistical characterization of the interferers around $x_{0}$ after this conditioning. % Clearly, the conditioning affects the activity status of the potential interferers as the transmission of $x_{0}$ may be sensed by the potential interferers. % in the sense that it introduces a bias to the question whether or not a potential interferer becomes active. 
%To overcome this analytic intractability the set of potential interferers is approximated by an non-homogeneous PPP in the second step. %Here, the problem of correctly capturing the interactions between the nodes surrounding $x_{0}$ and $x_{0}$ itself is virtually transformed into a location-dependent thinning of the PPP. 
2) This is where the non-homogeneous Poisson approximation comes into play, yielding a ``small-scale" density $\lambda_{s}$ 
\begin{IEEEeqnarray}{c}
	\lambda_{s}(r)=\lambda_{\ell}\left(1-\exp\left(-\Delta r^{\alpha}\right)\right),\label{eq:small_scale_density}
\end{IEEEeqnarray}
modeling the density of interferers around the reference transmitter $x_{0}$. The term $1-\exp\left(-\Delta r^{\alpha}\right)$ can be seen as the probability that an interferer at distance $r$ to $x_{0}$ does not sense the ongoing transmission of $x_{0}$.% The density $\lambda_{s}(r)$ is monotonically increasing in $r$ since it is more likely that far away potential interferers will be unable to sense the transmission of $x_{0}$.
The behavior of $\lambda_{s}(r)$ can be described by the spatial shape function of scenario d), cf. Fig.~\ref{fig:scenarios}. 3) The density $\lambda_{s}(r)$ in (\ref{eq:small_scale_density}) is then used to describe the interference \emph{around the reference receiver} $y_{0}$, although it reflects the interference experienced by the reference transmitter at $x_{0}$. This simplification step increases analytic tractability at the cost of losing accuracy. The level of accuracy loss resulting from step 3) is next studied for the Rayleigh fading model with $\alpha=4$ and $c=0$. We use the same notion as in (\ref{eq:rel_ap_error}) to measure the relative accuracy loss %\footnote{We use the term accuracy loss rather than approximation error in order to avoid misunderstandings: the two-step approach explained above for embedding the CSMA mechanism into the model already inherits an error due to the Poisson approximation. However, we are interested in the additional loss in accuracy that is induced by the simplification procedure described above.} 
as a function of $d$, i.e., 

\begin{figure}[!t]
	\psfrag{tagx}[c][c]{\small{$d$}}
	\psfrag{tagrel}[c][c]{\small{$\delta(d)$}}
	\psfrag{tagabs}[c][c]{\small{$|\op(o)-\op(y_{0})|$}}
	\psfrag{tag1ta:}{\small{$\beta=0.1$}}
	\psfrag{tag2}{\small{$\beta=1$}}
	\psfrag{tag3}{\small{$\beta=10$}}
	\psfrag{tag4tag4:}{\small{$\lambda=10^{-2}$}}
	\psfrag{tag5}{\small{$\lambda=10^{-3}$}}
	\psfrag{tag6ta}[c][c]{\small{$\beta=0.1$}}
	\psfrag{tag7}[c][c]{\small{$\beta=1$}}
	\psfrag{tag8}[c][c]{\small{$\beta=10$}}
	\centering
	\includegraphics[width=0.5\textwidth]{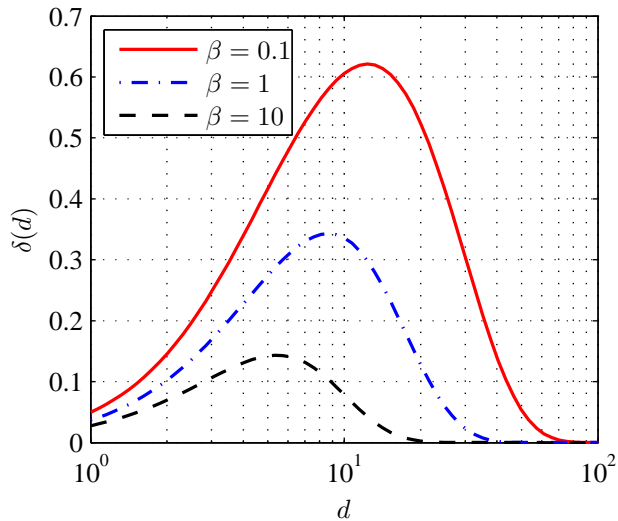}
	
	%\hfil
	%\subfloat[Absolute accuracy loss]{\includegraphics[width=0.5\textwidth]{../figures/figure_abs_error_csma_model.eps}
	%\label{fig:csma_abs_ap_er}}}
	\caption{(a) Relative accuracy loss $\delta(d)$ vs. transmission distance $d$ for different $\beta$. Parameters are $\alpha=4$, $c=0$, $\lambda=10^{-3}$ and $\Delta=-50$ dB.% (b) Absolute accuracy loss $|q(o)-q(y_{0})|$ vs. $d$ for different $\beta$ and $\lambda$. Parameters are $\alpha=4$, $c=0$. $\Delta$ was chosen such that the spatial throughput is maximized under the maximum outage probability constraint $\epsilon=0.25$.
}\label{fig:csma_rel_ap_er}
\end{figure}

\begin{IEEEeqnarray}{c}
	\delta(d)\triangleq\frac{|\op(o,d)-\op(y_{0},d)|}{\op(y_{0},d)},\quad \|y_{0}\|=d,\label{eq:def_rel_error_csma}
\end{IEEEeqnarray}
where the second argument in $\op$ now highlights the dependence on $d$. Fig.~\ref{fig:csma_rel_ap_er} shows the accuracy loss vs. $d$ for different $\beta$ and $\Delta=-50$ dB. The density of potential transmitters is $\lambda=10^{-3}$. It can be seen that depending on the value of $d$, significant errors can occur. %Unfortunately, they are spread over a wide range of $d$ with their highest value concentrated around $d=10$; a distance that is often used in the modeling of decentralized networks \cite{weber05,weber10}. 
Interestingly, the relative accuracy loss decreases with $\beta$ which is due to an increased outage correlation at the locations $x_{0}$ and $y_{0}$. % increases with $\beta$ which in turn increases their overlapping, resulting in a higher outage correlation. In other words, the sum interference created by nodes farther apart becomes more and more relevant as $\beta$ increases. In this case the interference field is no longer dominated by a few nearby nodes but it is dominated by many far away nodes that have a small relative difference between the path loss to $x_{0}$ and to $y_{0}$. On the opposite, at low $\beta$, e.g., when employing DS-CDMA, the interference is dominated by a few nearby interferers, thus rendering the interference experienced at $x_{0}$ and $y_{0}$ less correlated. %In addition, Fig.~\ref{fig:csma_abs_ap_er} shows the absolute accuracy loss in terms of outage probability deviation in order to characterize the QoS false-prediction. The parameter $\Delta$ was obtained by numerically optimizing spatial throughput subject to an outage probability constraint $\epsilon=0.25$ \cite{tanbourgi12_1}. The plot clearly demonstrates what was already discussed above: the accuracy loss increases with decreasing $\beta$ (error $>20\%$ for $\beta=0.1$). Lowering the density of potential interferers $\lambda$ effectively shifts high accuracy losses towards larger $d$. Note that assuming the reference receiver $y_{0}$ to be located in the origin yields a too \emph{optimistic} outage probability. In this example, this consequently means that in the dense regime, where $q(y_{0})=\epsilon$ (horizontal lines), the actual outage probability is $q(y_{0})-q(o)$ times higher than the targeted value of $\epsilon=0.25$ ($+21.6\%$ for $\beta=0.1$, $+9.6\%$ for $\beta=1$ and $+1.68\%$ for $\beta=10$).
The error remains small for small $d$, i.e., short-range communications, for the same reason as above. For very large $d$, $\delta(d)$ tends to zero which results from both $\op(o,d)$ and $\op(y_0,d)$ tending to zero.
%  
%In conclusion, we note that if the network is driven at high $\beta$, i.e., spectral efficiency of transmission is high, the accuracy loss induced by assuming that the reference receiver sees exactly the same interference field as the reference transmitter may be negligible. Unless the transmission distance $d$ is ultimately small, i.e., short-range transmission, the exact location of the reference receiver $y_{0}$ relative to the reference transmitter $x_{0}$ must be considered in all other cases in order to avoid high losses in accuracy. 

\section{Conclusion}\label{sec:conclusion}
We extended prior work on interference modeling for wireless networks with isotropic but not necessarily stationary spatial distribution of nodes. The interference statistics and the outage probability were analyzed as a function of (i) an arbitrary receiver location inside the network and (ii) an arbitrary but isotropic node distribution. For the path loss exponents $\alpha=2$ and $\alpha=4$ closed-form expressions were obtained while bounds were derived for other cases. The developed model led to some interesting insights that could not have been revealed previously due to limiting the analysis to stationary models only. % not  not a For instance, for $\alpha=2$, there exists a transition between sparse and dense networks where interference for instance  it is possible that the interference is almost surely finite while the number of contributing interferers is almost surely infinite; thereby demonstrating the existence of a sparse and a dense regime. This particular result sharpens prior statements on the interference characterization for the case $\alpha=2$. The interference analysis was conducted for arbitrary fading models (including the pure path loss case) as well as for the Rayleigh fading model. Moreover, we presented a lower and upper bound on the tail probability of the interference for arbitrary fading, where the lower bound is not limited to $\alpha=2$ and $\alpha=4$. 
The usefulness of the results was discussed and demonstrated through examples in network analysis related to outage probability, local throughput characterization and carrier-sensing mechanisms. It was found that the developed model increases model accuracy significantly and provides an adequate tool to describe location-depended performance in networks with practical node distribution.%transmitter-inhibition Furthermore, we proposed two performance metrics, namely the local transmission capacity and the sum spatial throughput, which are suitable for measuring the local throughput in such non-stationary networks. The local transmission capacity is a refinement of the well-known transmission capacity, which fails to characterize throughput in networks with non-stationary spatial node distributions. This refinement allows for comparing and benchmarking different protocols over the network area, and provides spatial information about the local throughput. The sum spatial throughput counts the average number of successful transmissions in the network, thereby accounting for boundary effects and other non-homogeneities in the spatial distribution of nodes. The sum spatial throughput may be of special interest for optimizing finite networks, which due to their non-stationary nature, cannot be properly described using the traditional spatial throughput metric.

%We applied our interference model developed in this work to these two metrics and discussed their importance with respect to some problems in network design. For example, using the local transmission capacity we were able to show that for very small path loss exponents ($\alpha\to2$) the throughput gain of FH-CDMA over DS-CDMA scales logarithmically with the processing gain, and thus is non-vanishing as opposed to prior investigations. Our model recovers accuracy losses coming from local approximations that  neglect boundary effects and/or non-homogeneous deployments in existing networks. Seeing the model as a design tool, it may be of significant importance for the correct dimensioning of system parameters in order to satisfy QoS requirements. We also showed that the developed model can also help to better describe networks that employ CSMA sensing at the transmitter.



\appendices

\section{Integral Identities}\label{sec:integrals}
\begin{identity}\label{lemma:integral_id1}
	 If $a>|b|$, $a,b\in\mathbb{R}$, than
	    \begin{IEEEeqnarray}{c}
			  \int_{0}^{\pi}\frac{\mathrm d\phi}{(a+b\cos\phi)^{n+1}}=\frac{\pi\,P_{n}\left(\frac{a}{\sqrt{a^2-b^2}}\right)}{(a^2-b^2)^{\frac{n+1}{2}}},\IEEEeqnarraynumspace\label{eq:integral_id1}
	    \end{IEEEeqnarray}
	    where $P_{n}(x)$ is the $n^{\text{th}}$-Legendre polynomial \cite{gradshteyn07}. We will be using $n=1$, leading to the $0^{\text{th}}$-Legendre polynomial given by $P_{0}(x)=1$.  
\end{identity}
%\begin{IEEEproof}
%	This correspondence can be found in \cite{gradshteyn07}.
%\end{IEEEproof}
\begin{identity}\label{lemma:integral_id2}
	Let $a_{1},a_{2},a_{3}\in\mathbb{R}$, where $a_{3}>0$. Define $R\triangleq a_{1}+a_{2}t^2+a_{3}t^4$, $T=4a_{1}a_{3}-a_{2}^2$. By \cite{gradshteyn07} and using the substitution $t\to t^2$, we have
	\begin{IEEEeqnarray}{c}
		\int\frac{2t\sqrt{a_{3}}}{\sqrt{a_{1}+a_{2}t^2+a_{3}t^4}}\,\mathrm dt=\begin{cases}
				\log\frac{2\sqrt{a_{3}R}+2a_{3}t^2+a_{2}}{\sqrt{T}},& a_{3}>0\\
		            \text{\textnormal{asinh}}\frac{2a_{3}t^2+a_{2}}{\sqrt{T}}, & T>0\\
		            \log(a_{3}t^2+\tfrac{a_{2}}{2}), & T=0.
		            \end{cases}\IEEEeqnarraynumspace\label{eq:integral_id2}
	\end{IEEEeqnarray}
\end{identity}

\begin{identity}\label{lemma:integral_id3}
	Let $a_{1},a_{2}\in\mathbb{R}$, where $a_{1}>0$. Then,
	\begin{IEEEeqnarray}{c}
	\int\int_{0}^{\pi}\frac{2t}{a_{1}+(t^2+a_{2}^2-2ta_{2}\cos\phi)^2}\,\mathrm d\phi\,\mathrm d t=\frac{\pi}{2\sqrt{a_{1}}}\text{\textnormal{atan}}\frac{2\mathfrak{R}\{\kappa(t,a_{1},a_{2})\}}{1-|\kappa(t,a_{1},a_{2})|^2},\IEEEeqnarraynumspace\label{eq:pbz}
	\end{IEEEeqnarray}
	where
	\begin{IEEEeqnarray}{c}
		\kappa(t,a_{1},a_{2})\triangleq\frac{t^2-a_{2}^2-j\sqrt{a_{1}}}{\sqrt{(\sqrt{a_{1}}+j(t^2+a_{2}^2))^2+4t^2a_{2}^2}}.\label{eq:kappa}\IEEEeqnarraynumspace
	\end{IEEEeqnarray}
\end{identity}
%Identity 3 deserves a proof:
\begin{IEEEproof}
	The basic idea is to decompose the integrand into partial fractions and to apply Identity \ref{lemma:integral_id1} and \ref{lemma:integral_id2}, yielding (\ref{eq:pbz})  after some algebraic manipulations. Note that according to \cite{gradshteyn07}, (\ref{eq:integral_id1}) and (\ref{eq:integral_id2}) hold only for real-valued parameters. However, we verified that they also hold for complex-valued parameters. %Thus, we have
	%\begin{IEEEeqnarray}{c}
	%	\frac{1}{2\sqrt{c}}\sum\limits_{k=-1,1}\int t \int_{0}^{\pi}\frac{\mathrm d\phi \,\mathrm dt}{\sqrt{a_{1}}+kj(t^2+a_{2}^2-2ta_{2}\cos\phi)},\IEEEnonumber
	%\end{IEEEeqnarray}
	%where the inner integral can be calculated using Lemma \ref{lemma:integral_id1}. The result then follows after some algebraic manipulations and integration. 
\end{IEEEproof}

\section{Proofs}\label{sec:proofs}

\subsection{Proof of Theorem \ref{theorem:moment1}}\label{proof:moment1}
We want to compute the expectation
	\begin{IEEEeqnarray}{c}
		\mathbb{E}\left[\sum\limits_{\mathsf{x}\in\Phi}\mathsf{g}_{\mathsf{x}}\ell(\|\mathsf{x}-y_{0}\|)\right],
	\end{IEEEeqnarray}
	where the expectation is with respect to the interferer locations and the channel gains $\mathsf{g}_{\mathsf{x}}$. Since the expectation operator linear, we can compute the expectation with respect to all $\mathsf{g}_{\mathsf{x}}$ first, i.e., $\mathbb{E}\left[\mathsf{g}_{\mathsf{x}}\right]=1$ $\forall\,\mathsf{x}\in\Phi$. %We then apply Slivnyak's Theorem, which states that the law of the PPP $\Phi$ is not changed when $\Phi$ is conditioning on having a fixed point at $x_{0}$, i.e., $\mathbb{E}\left[\mathsf{I}(y_{0})\right]=\mathbb{E}\left[\mathsf{I}(y_{0})\right]$, yielding
	%\begin{IEEEeqnarray}{c}
	%	\mathbb{E}\left[\mathsf{I}(y_{0})\right]=\mathbb{E}\left[\sum\limits_{\mathsf{x}\in\Phi}\ell(|\mathsf{x}-y_{0}|)\right].
	%\end{IEEEeqnarray}
	Applying the Campbell Theorem, yields
	\begin{IEEEeqnarray}{c}
		\mathbb{E}\left[\mathsf{I}(y_{0})\right]= \lambda\int_{\mathbb{R}^2}\ell(\|x-y_{0}\|) F(\|x\|) \,\mathrm dx.\label{eq:proof_moment1_1}
	\end{IEEEeqnarray}
	Changing to polar coordinates and exploiting the isotropy property, we can rewrite (\ref{eq:proof_moment1_1}) as
	\begin{IEEEeqnarray}{c}
		\mathbb{E}\left[\mathsf{I}(y_{0})\right]= \lambda\int_{0}^{\infty}\int_{0}^{\pi}\frac{2r\,F(r)}{c+(r^2+\|y_{0}\|^2-2r\|y_{0}\|\cos\phi)}\,\mathrm d\phi\,\mathrm dr.\label{eq:proof_moment1_2}\IEEEeqnarraynumspace
	\end{IEEEeqnarray}
	We then apply Identity \ref{lemma:integral_id1} to the inner integral of (\ref{eq:proof_moment1_2}) to obtain
	\begin{IEEEeqnarray}{c}
		\mathbb{E}\left[\mathsf{I}(y_{0})\right]= \lambda\pi\int_{0}^{\infty}\frac{2r\,F(r)}{\sqrt{ (c+r^2+\|y_{0}\|^2)^2-4r^2\|y_{0}\|^2}}\,\mathrm dr.\label{eq:proof_moment1_3}\IEEEeqnarraynumspace
	\end{IEEEeqnarray}
	Finally using product integration and applying Identity \ref{lemma:integral_id2} to (\ref{eq:proof_moment1_3}) and verifying the convergence of the upper limit using the constraint $ F(r)\widesim{r\to\infty}\frac{1}{r^\nu}$ for some $\nu>0$, yields the result.\qed
	
\subsection{Proof of Theorem \ref{thm:dense_network}}\label{proof:dense_network}
To prove that $\mathsf{I}(y_{0})$ is a.s. infinite, we analyze its Laplace transform $\mathbb{E}\left[\exp(-s\mathsf{I}(y_{0}))\right]$ and check if it is zero for all $s$. In the PPP case, the Laplace transform of the interference field is given by \cite{baccelli09a,ganti09}
\begin{IEEEeqnarray}{c}
	\mathbb{E}\left[\exp(-s\mathsf{I}(y_{0}))\right]=\exp\left(-\mathbb{E}_{\mathsf{g}}\left[\int_{\mathbb{R}^2}\left(1-e^{-s\mathsf{g}\ell(\|x-y_{0}\|)}\right)\,\lambda(x)\,\mathrm dx\right]\right).\label{eq:laplace_int}\IEEEeqnarraynumspace
\end{IEEEeqnarray}
For (\ref{eq:laplace_int}) to become zero, the integral must not converge. We write
\begin{IEEEeqnarray}{c}
	\int_{\mathbb{R}^2}\left(1-e^{-s\mathsf{g}\ell(\|x-y_{0}\|)}\right)\,\lambda(x)\,\mathrm dx\overset{(a)}{\geq}\int_{\mathbb{R}^2}\frac{s\mathsf{g}\ell(\|x-y_{0}\|)}{1+s\mathsf{g}\ell(\|x-y_{0}\|)}\,\lambda(x)\,\mathrm dx,\label{eq:laplace_int2}\IEEEeqnarraynumspace
\end{IEEEeqnarray}
where (a) follows from the inequality $\frac{z}{1+z}\leq 1-e^{-z}$ for $z>-1$ \cite{olver10}. Inserting the path loss function $\ell(\|x-y_{0}\|)=(c+\|x-y_{0}\|^{2})^{-1}$ into the right-hand side of (\ref{eq:laplace_int2}) yields
\begin{IEEEeqnarray}{c}
	2\lambda s\mathsf{g}\int_{0}^{\pi}\int_{0}^{\infty}\frac{rF(r)}{s\mathsf{g}+c+r^2+y_{0}^2-2ry_{0}\cos\phi}\,\mathrm dr\,\mathrm d\phi.
\end{IEEEeqnarray}
At the upper limit of the inner integral the integrand behaves as $F(r)/r$. So, the condition $0<\lim_{r\to\infty}F(r)r^{\nu}<\infty$, where $\nu\to0$, is sufficient for the divergence of the integral. Because the Laplace transform of $\mathsf{I}(y_{0})$ becomes zero in this case, this concludes the proof.\qed

\subsection{Proof of Theorem \ref{theorem:moment2}}\label{proof:moment2}
The proof is analogous to the proof of Theorem \ref{theorem:moment1}, except for the integration part. Here, the integral Identity \ref{lemma:integral_id3} is used instead. We further exploit the fact that $\max_{r}\{F(r)\}=1$ to ensure convergence of the integrals.\qed

\begin{figure}[t]
	\psfrag{a}[c][c]{\small{$y_{0}$}}
	\psfrag{b}[c][c][1][55]{\small{$F(\|x\|)$}}
	\psfrag{c}{\small{$\bar r(y_{0})$}}
		   \centering
	\includegraphics[width=0.5\textwidth]{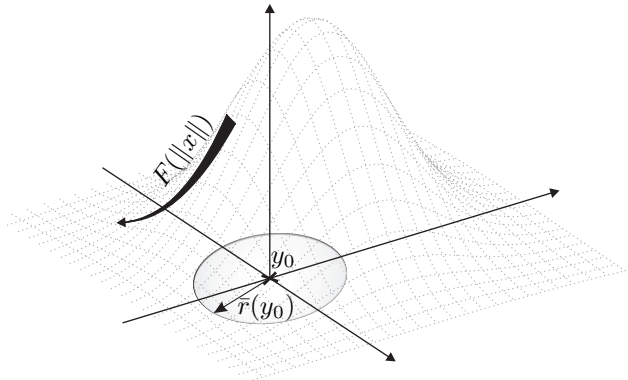}
\caption{Schematic illustration of the proof of Theorem \ref{theorem:lower_bound}. Shaded area around $y_{0}$ represents the largest disc contained in the subharmonic region $G$.}% In case of pure path loss model ($\mathsf{g}\equiv1$), the radius of this disc is equal to $\min\{\bar r^2(y_{0}),(\tfrac{1}{z}-c)^{\frac{2}{\alpha}}\}$.}
\label{fig:proof_subharmonic}
\end{figure} 

\subsection{Proof of Theorem \ref{theorem:lower_bound}}\label{proof:lower_bound}
We follow the idea of dominant interferers which was introduced in \cite{weber05} to bound the tail probability $\mathbb{P}(\mathsf{I}(y_{0})\geq z)$: let $\mathcal{A}(z)\triangleq\{(\mathsf{x},\mathsf{g}_{\mathsf{x}})\in\Phi\times\mathbb{R}_{+}:\mathsf{g}_{\mathsf{x}}\ell(\|\mathsf{x}-y_{0}\|)\geq z\}$ denote the set of all interferers, where each one taken by itself already results in the event $\mathsf{I}(y_{0})\geq z$. Clearly,
\begin{IEEEeqnarray}{c}	\mathsf{I}(y_{0})=\sum_{(\mathsf{x},\mathsf{g}_{\mathsf{x}})\in\mathcal{A}(z)}\mathsf{g}_{\mathsf{x}}\ell(\|\mathsf{x}-y_{0}\|)+\sum_{(\mathsf{x},\mathsf{g}_{\mathsf{x}})\in\mathcal{\bar A}(z)}\mathsf{g}_{\mathsf{x}}\ell(\|\mathsf{x}-y_{0}\|),\IEEEeqnarraynumspace\label{eq:dom_interference}
\end{IEEEeqnarray}
where the set $\mathcal{\bar A}$ is denotes the elements outside $\mathcal{A}$ on the same domain, i.e., $\mathcal{\bar A}(z)\triangleq\{(\mathsf{x},\mathsf{g}_{\mathsf{x}})\in\Phi\times\mathbb{R}_{+}:\mathsf{g}_{\mathsf{x}}\ell(\|\mathsf{x}-y_{0}\|)< z\}$. Note that both sums in (\ref{eq:dom_interference}) are non-negative. Hence, we can now write
\begin{IEEEeqnarray}{rCl}
	\mathbb{P}(\mathsf{I}(y_{0})\geq z)&\overset{(a)}{\geq}& \mathbb{P}\left(\sum_{(\mathsf{x},\mathsf{g}_{\mathsf{x}})\in\mathcal{A}(z)}\mathsf{g}_{\mathsf{x}}\ell(\|\mathsf{x}-y_{0}\|)\geq z\right)\IEEEnonumber\IEEEeqnarraynumspace\\
		&\overset{(b)}{=}& \mathbb{P}\left(|\mathcal{A}(z)|>0\right)\IEEEnonumber\\
		&\overset{(c)}{=}& 1-\exp\left(-\Lambda(\mathcal{A}(z))\right),\label{eq:lower_bound_poisson}
\end{IEEEeqnarray}
where (a) follows from removing the ``non-dominant'' part in (\ref{eq:dom_interference}), (b) follows from the definition of $\mathcal{A}(z)$ and (c) is a consequence of $|\mathcal{A}(z)|$ being Poisson distributed with mean $\Lambda(\mathcal{A}(z))$. Using \cite[Corollary 2.1.2]{baccelli09a}, $\Lambda(\mathcal{A}(z))$ in (\ref{eq:lower_bound_poisson}) can be computed as
\begin{IEEEeqnarray}{c}
	\Lambda(\mathcal{A}(z)) = \int_{\mathbb{R}^2}\mathbb{P}\left(\mathsf{g}\geq z\,(c+\|x-y_{0}\|^{\alpha})\right) \lambda(x)\,\mathrm dx. \IEEEeqnarraynumspace
\end{IEEEeqnarray}
We now translate $\Phi$ by the vector $y_{0}$ to obtain a $y_{0}$-centric coordinate system, cf. Fig. \ref{fig:proof_subharmonic}. %Furthermore, we insert the spatial shape function $F(x)$ that we have re-defined for the two-dimensional case. 
After switching to polar coordinates we obtain
\begin{IEEEeqnarray}{c}
	\Lambda(\mathcal{A}(z)) = \int_{0}^{\infty}r\mathbb{P}\left(\mathsf{g}\geq z\,(c+r^{\alpha})\right)\int_{0}^{2\pi} \lambda(y_{0}+r e^{j\phi})\,\mathrm d\phi\,\mathrm dr.\IEEEeqnarraynumspace\label{eq:lower_bound_poisson2}
\end{IEEEeqnarray}
We now exploit the fact that $\lambda(x)$ is subharmonic around $y_{0}$ in a region $G$. Let $\bar r(y_{0})$ denote the maximal radius for which $b(y_{0},\bar r(y_{0}))$ is contained in $G$. Then, (\ref{eq:lower_bound_poisson2}) can be bounded as  
\begin{IEEEeqnarray}{rCl}
	\Lambda(\mathcal{A}(z))&\overset{(a)}{\geq}&\int_{0}^{\bar r(y_{0})}r\,\mathbb{P}\left(\mathsf{g}\geq z\,(c+r^{\alpha})\right)\int_{0}^{2\pi} \lambda(y_{0}+r e^{j\phi})\,\mathrm d\phi\,\mathrm dr\IEEEnonumber\\
	&\overset{(b)}{\geq}& 2\pi \lambda(y_{0})\int_{0}^{\bar r(y_{0})}r\,\mathbb{P}\left(\mathsf{g}\geq z\,(c+r^{\alpha})\right)\,\mathrm dr\IEEEnonumber\\
	&\overset{(c)}{=}& 2\pi\lambda F(\|y_{0}\|)\int_{0}^{\bar r(y_{0})}r\,\mathbb{P}\left(\mathsf{g}\geq z\,(c+r^{\alpha})\right)\,\mathrm dr,\IEEEeqnarraynumspace\label{eq:subharmonicity}
\end{IEEEeqnarray}
where (a) follows from limiting the upper integration limit to $\bar r(y_{0})$. Inequality (b) is a consequence of subharmonicity \cite[Ch.10]{conway86}, and (c) follows from Definition 1. Inserting (\ref{eq:subharmonicity}) in (\ref{eq:lower_bound_poisson}) yields the result.\qed

\subsection{Proof of Theorem \ref{theorem:laplace_int}}
	We write
	\begin{IEEEeqnarray}{rCl}
	\mathcal{L}_{\mathsf{I}(y_{0})}(s)&\overset{(a)}{=}&\mathbb{E}_{\Phi}\left[\prod\limits_{\mathsf{x}\in\Phi} \mathbb{E}_{\mathsf{g}_{\mathsf{x}}}\Big[\exp\left(-s\mathsf{g}_{\mathsf{x}}\ell(\|\mathsf{x}-y_{0}\|)\right)\Big]\right]\IEEEnonumber\\
		&\overset{(b)}{=}&\exp\left(-\int_{\mathbb{R}^2}\Big(1-\mathcal{L}_{\mathsf{g}}\left(s\ell(\|x-y_{0}\|)\right)\Big)\,\lambda(x)\,\mathrm dx\right),\IEEEnonumber\IEEEeqnarraynumspace
	\end{IEEEeqnarray}
	where (a) follows from algebraic manipulations and the i.i.d. property of the $\mathsf{g}_{\mathsf{x}}$. (b) follows from the probability generating functional and the Laplace functional of a PPP \cite{baccelli09a}. After noting that $\mathcal{L}_{\mathsf{g}}(s)=(1+s)^{-1}$ for exponentially distributed $\mathsf{g}$, the integral is computed using Identity \ref{lemma:integral_id1} and Identity \ref{lemma:integral_id2} for $\alpha=2$ and Identity \ref{lemma:integral_id3} for $\alpha=4$. %\cite{guo13}

Note that (a) in the proof holds for general point processes and some approximation techniques for computing the right-hand side already exist \cite{ganti10}. The (b) part is for PPPs only.\qed

% \subsection{Proof of Corollary \ref{col:op_rayleigh}}
% It is well-known that the OP for Aloha MAC and exponentially distributed power gains $\mathsf{g}_{x}$ can be written in terms of the Laplace transform of the interference \cite{baccelli09,ganti09}: We condition (\ref{eq:sinr}) on $\Phi$ and evaluate the OP first with respect to $\mathsf{g}_{x_{0}}$. We finally use (\ref{eq:laplace}) with $s=\beta(c+d^{\alpha})$.\qed

\subsection{Proof of Corollary \ref{col:fh_ds_gain}}
Solving (\ref{eq:op}) for $\lambda$ and multiplying by $1-\op(y_{0})$ yields the local transmission capacity
	\begin{IEEEeqnarray}{c}
		c(y_{0},\epsilon)=\frac{-\log(1-\epsilon)(1-\epsilon)}{\beta d^2 A_{2}(y_{0},\beta d^2)}\label{eq:proof_gain_FH_DS1}
	\end{IEEEeqnarray}
	after substituting $\op(y_{0})\to\epsilon$. We are interested in the case $y_{0}=o$. Using Identity 2 (case $T=0$) we find that
	\begin{IEEEeqnarray}{c}
		A_{2}(o,\beta d^2)=-F(0)\log\beta d^2-\int_{0}^{\infty}f(r)\log(r^2+\beta d^2)\,\mathrm dr.\IEEEeqnarraynumspace\label{eq:proof_gain_FH_DS2}
	\end{IEEEeqnarray}
	Assuming that nodes employ pseudo-noise sequences, FH effectively thins out the point process of interferers by $M$ (interference avoidance, $\lambda/M$), while DS reduces the interference by a factor of $M$ (interference averaging, $\beta/M$), cf. \cite{weber05}. Hence, using (\ref{eq:proof_gain_FH_DS1}) and (\ref{eq:proof_gain_FH_DS2}) the ratio $\tfrac{c^{\text{FH}}(0,\epsilon)}{c^{\text{DS}}(0,\epsilon)}$ can be written as
	\begin{IEEEeqnarray}{rCl}
		\frac{c^{\text{FH}}(o,\epsilon)}{c^{\text{DS}}(o,\epsilon)}&=&\frac{M}{\beta d^2 A_{2}(o,\beta d^2)}\frac{\beta d^2}{M}A_{2}(o,\tfrac{\beta}{M} d^2)\IEEEnonumber\\
		&=&\frac{A_{2}\left(0,\frac{\beta}{M} d^2\right)}{A_{2}(0,\beta d^2)}\IEEEnonumber\\
		&\overset{(a)}{=}&\frac{-F(0)\log\frac{\beta}{M} d^2}{A_{2}(o,\beta d^2)}-\frac{\int_{0}^{\infty}f(r)\log\left(r^2+\frac{\beta}{M} d^2\right)\,\mathrm dr}{A_{2}(o,\beta d^2)}\IEEEnonumber\\
		%&=&\frac{F(0)\log M-F(0)\log\beta d^2}{A_{2}(o,\beta d^2)}-\frac{\int_{0}^{\infty}f(r)\log\left(r^2+\frac{\beta}{M} d^2\right)\,\mathrm dr}{A_{2}(o,\beta d^2)}\IEEEnonumber\\
		&\overset{(b)}{=}&1+\frac{F(0)\log M}{A_{2}(o,\beta d^2)}+\frac{1}{A_{2}(o,\beta d^2)}\int_{0}^{\infty}f(r)\log\frac{r^2+\beta d^2}{r^2+\tfrac{\beta}{M} d^2}\,\mathrm dr,\IEEEeqnarraynumspace\label{eq:proof_gain_FH_DS3}
	\end{IEEEeqnarray}
	where (a) follows from (\ref{eq:op}) and Remark \ref{rem:a2} and (b) follows from algebraic manipulations. Now we show that the integral in (\ref{eq:proof_gain_FH_DS3}) is finite. Assuming $|f(r)|<\infty$ (which is reasonable in practical networks), we note that the integrand has no singular values.
%\begin{IEEEeqnarray}{rCl}
%	&&\int_{0}^{\infty}f(r)\left(\log(r^2+\beta d^2)-\log(r^2+\tfrac{\beta}{M} d^2)\right)\,\mathrm dr\IEEEnonumber\\
%	&&\hspace{0.2cm}=\int_{0}^{\infty}f(r)\log\frac{r^2+\beta d^2}{r^2+\tfrac{\beta}{M} d^2}\,\mathrm dr\IEEEeqnarraynumspace
	%&=&-\int_{0}^{\infty}f(r)\log\frac{r^2+\beta d^2+\frac{\beta}{M} d^2-\beta d^2}{r^2+\beta d^2}\,\mathrm dr\IEEEeqnarraynumspace\\
	%&=&-\int_{0}^{\infty}f(r)\log1+\frac{\frac{\beta}{M} d^2-\beta d^2}{r^2+\beta d^2}\,\mathrm dr\IEEEeqnarraynumspace\\
	%&\leq&-\int_{0}^{\infty}f(r)\log1+\frac{|\frac{\beta}{M} d^2-\beta d^2|}{r^2+\beta d^2}\,\mathrm dr\IEEEeqnarraynumspace\\
	%&\leq&-|\frac{1}{M} -1|\int_{0}^{\infty}f(r)\frac{\beta d^2}{r^2+\beta d^2}\,\mathrm dr\IEEEeqnarraynumspace=o(1)
%\end{IEEEeqnarray}
For $M=1$ the logarithm becomes zero and so does the integral. Since $\log\frac{r^2+\beta d^2}{r^2+\frac{\beta d^2}{M}}\leq\log(1+\frac{\beta d^2}{r^{2}})$ for all $M>0$, we therefore analyze the convergence of the integral
\begin{IEEEeqnarray}{c}
	\Big|\int_{0}^{\infty}f(r)\log(1+\tfrac{\beta d^2}{r^{2}})\,\mathrm dr\Big|\leq\underset{r}{\max}\{|f(r)|\}\int_{0}^{\infty}\log(1+\tfrac{\beta d^2}{r^{2}})\,\mathrm dr<\infty.\IEEEeqnarraynumspace
\end{IEEEeqnarray}
Hence, the integral in (\ref{eq:proof_gain_FH_DS3}) is finite.\qed

\bibliographystyle{IEEEtran}
\bibliography{IEEEabrv,../../literature}

\end{document}